\documentclass[a4]{article}
\usepackage{arxiv}

\usepackage[utf8]{inputenc} 
\usepackage[T1]{fontenc}    

\usepackage[
    backend=biber,
    style=numeric-comp,
    minnames=5,
    maxnames=5,
    backref=true
]{biblatex}
\usepackage{placeins}
\usepackage{hyperref}       
\usepackage{url}            
\usepackage{booktabs}       
\usepackage{amsfonts}       
\usepackage{nicefrac}       
\usepackage{microtype}      
\usepackage{lipsum}
\usepackage{graphicx}

\usepackage{color}
\usepackage{xspace}
\usepackage{amsmath,amssymb,amsthm}
\usepackage{csquotes}
\usepackage{booktabs}
\usepackage{enumitem}
\usepackage[nameinlink,capitalize,noabbrev]{cleveref}
\usepackage{tabularx}
\usepackage{orcidlink}

\newcommand{\citecmd}[1]{%
~\cite{#1}%
}

\newcommand{\captionwithtitle}[2]{\caption{\textbf{#1}. #2}}

\newcommand{\qquote}[1]{``#1''}

\newcolumntype{Y}{>{\centering\arraybackslash}X}

\title{SALT: Introducing a Framework for Hierarchical Segmentations in Medical Imaging using Softmax for Arbitrary Label Trees}
\makeatletter
\newcommand{\myfnsymbol}[1]{%
  \expandafter\@myfnsymbol\csname c@#1\endcsname
}

\newcommand{\@myfnsymbol}[1]{%
  \ifcase #1
  \or 1
  \or 2
  \or 3
  \or 4
  \or 5
  \or \TextOrMath{\textasteriskcentered}{*}
  \or \TextOrMath{\textdagger}{\dagger}
  \fi
}
\newcommand{\radiology}{\@myfnsymbol{1}}
\newcommand{\ikim}{\@myfnsymbol{2}}
\newcommand{\transfusion}{\@myfnsymbol{3}}
\newcommand{\imibe}{\@myfnsymbol{4}}
\newcommand{\fh}{\@myfnsymbol{5}}
\newcommand{\equalcontributor}{\@myfnsymbol{6}}
\newcommand{\correspondingA}{\@myfnsymbol{7}}

\makeatother

\author{
Sven Koitka \textsuperscript{\radiology,\ikim,\equalcontributor}%
\orcidlink{0000-0001-9704-1180}\\
\And Giulia Baldini \textsuperscript{\radiology,\ikim,\equalcontributor}%
\orcidlink{0000-0002-5929-0271}\\
\And Cynthia S. Schmidt \textsuperscript{\ikim,\transfusion}%
\orcidlink{0000-0003-1994-0687}\\
\And Olivia B. Pollok \textsuperscript{\ikim}\\
\And Obioma Pelka \textsuperscript{\ikim}\\
\And Judith Kohnke \textsuperscript{\radiology,\ikim}\\
\And Katarzyna Borys
\textsuperscript{\radiology,\ikim}%
\orcidlink{0000-0001-6987-6041}\\
\And Christoph M. Friedrich \textsuperscript{\imibe,\fh}%
\orcidlink{0000-0001-7906-0038}\\
\And Benedikt M. Schaarschmidt \textsuperscript{\radiology}%
\orcidlink{0000-0001-6331-7679}\\
\And Michael Forsting \textsuperscript{\radiology}\\
\And Lale Umutlu \textsuperscript{\radiology}\\
\And Johannes Haubold \textsuperscript{\radiology,\ikim}%
\orcidlink{0000-0003-4843-5911}\\
\And Felix Nensa \textsuperscript{\radiology,\ikim}%
\orcidlink{0000-0002-5811-7100}\\
\And René Hosch \textsuperscript{\radiology,\ikim,\correspondingA}%
\orcidlink{0000-0003-1760-2342}
}

\begin{document}

\maketitle

\renewcommand{\thefootnote}{\myfnsymbol{footnote}}
\maketitle
\footnotetext[1]{Institute of Interventional and Diagnostic Radiology and Neuroradiology, University Hospital Essen, Essen, Germany}%
\footnotetext[2]{Institute for Artificial Intelligence in Medicine (IKIM), University Hospital Essen, Essen, Germany}%
\footnotetext[3]{Institute for Transfusion Medicine, University Hospital Essen, Essen, Germany}%
\footnotetext[4]{Institute for Medical Informatics, Biometry and Epidemiology (IMIBE), University Hospital Essen, Essen, Germany}%
\footnotetext[5]{Faculty of Computer Science, University of Applied Sciences and Arts, Dortmund, Germany}%
\footnotetext[6]{Sven Koitka and Giulia Baldini contributed equally}%
\footnotetext[7]{Corresponding author: Rene.Hosch@uk-essen.de}%

\setcounter{footnote}{0}
\renewcommand{\thefootnote}{\fnsymbol{footnote}}

\begin{abstract}
\textbf{Background}:

Traditional segmentation networks approach anatomical structures as standalone elements, overlooking the intrinsic hierarchical connections among them. This study introduces Softmax for Arbitrary Label Trees (SALT), a novel approach designed to leverage the hierarchical relationships between labels, improving the efficiency and interpretability of the segmentations.

\textbf{Materials and Methods}:

This study introduces a novel segmentation technique for CT imaging, which leverages conditional probabilities to map the hierarchical structure of anatomical landmarks, such as the spine's division into lumbar, thoracic, and cervical regions and further into individual vertebrae. The model was developed using the SAROS dataset from The Cancer Imaging Archive (TCIA), comprising 900 body region segmentations from 883 patients. The dataset was further enhanced by generating additional segmentations with the TotalSegmentator, for a total of 113 labels. The model was trained on 600 scans, while validation and testing were conducted on 150 CT scans. Performance was assessed using the Dice score across various datasets, including SAROS, CT-ORG, FLARE22, LCTSC, LUNA16, and WORD. Additionally, 95\% confidence intervals (CI) were computed using 1000 rounds of bootstrapping.

\textbf{Results}:

Among the evaluated datasets, SALT achieved its best results on the LUNA16 and SAROS datasets, with Dice scores of 0.93 (95\% CI: 0.919, 0.938) and 0.929 (95\% CI: 0.924, 0.933) respectively. Additionally, the model demonstrated reliable accuracy across other datasets, scoring 0.891 (95\% CI: 0.869, 0.906) on CT-ORG and 0.849 (95\% CI: 0.844, 0.854) on FLARE22. Moreover, the LCTSC dataset showed a score of 0.908 (95\% CI: 0.902, 0.914) and the WORD dataset also showed good performance with a score of 0.844 (95\% CI: 0.839, 0.85). Furthermore, SALT is capable of segmenting a whole-body CT with 1000 slices in an average of 35 seconds.

\textbf{Conclusion}:

SALT used the hierarchical structures inherent in the human body to achieve whole-body segmentations with an average of 35 seconds per CT scan.  This rapid processing underscores its potential for integration into clinical workflows, facilitating the automatic and efficient computation of full-body segmentations with each CT scan, thus enhancing diagnostic processes and patient care.
\end{abstract}

\section{Introduction}
\label{introduction}

Computed Tomography (CT) imaging stands out as one of the most comprehensive\citecmd{junn_imaging_2021,neubauer_diagnostic_2023,osborne-grinter_prevalence_2023,guimaraes_advancements_2023} tools in the field of diagnostic imaging, with the number of CT scans increasing by 4\% per year worldwide\citecmd{schockel_developments_2020}. With the rising volume of CT scans, radiologists also face a growing workload, highlighting the importance of adopting automated solutions for support. Moreover, CT scans contain a substantial volume of information, and only a portion is used for specific diagnosis purposes. A substantial amount of potentially clinically valuable information remains unexploited and is the focus of current research\citecmd{jiang_radiomics_2022,bodden_incidental_2023, wasserthal_totalsegmentator_2023}. In this context, deep learning networks have the capability of automating tasks and extracting information from scans with little additional cost besides algorithm training. In particular, automated segmentation of CT scans is now a widespread technique for identifying key anatomical features such as organs, tissues, and vessels\citecmd{wasserthal_totalsegmentator_2023,koitka_fully_2022,radiya_performance_2023}. These detailed segmentations aid radiologists in making accurate diagnoses\citecmd{aromiwura_artificial_2023} and have been linked to indicators of a patient's well-being\citecmd{jung_association_2023,alderuccio_quantitative_2023}. This capability enables automated quantification of segmentations, with metrics such as organ volumetries gaining recognition for their role in predicting overall survival outcomes\citecmd{ito_spleen_2023,khoshpouri_quantitative_2019}. Furthermore, CT scans enable the automated calculation of Body Composition Analysis (BCA), which quantifies the amount of fat, muscle, and bone\citecmd{koitka_fully_2021,haubold_boa_2024,nowak_fully_2020,nowak_end--end_2022} and is proving to be valuable in monitoring disease progression\citecmd{jung_association_2023,mason_respiratory_2021,ko_change_2022} and predicting patient survival outcomes\citecmd{hosch_biomarkers_2022,keyl_deep_2023}.

The currently existing models for full-body segmentation, such as TotalSegmentator\citecmd{wasserthal_totalsegmentator_2023,isensee_nnu-net_2021}, often rely on multiple models for segmenting CT scans when the number of labels becomes too extensive for a single model to handle efficiently. Additionally, completing a full segmentation can take several minutes, which is impractical for scenarios where a segmentation algorithm is expected to constantly operate in the background as part of a hospital's data acquisition process.

In response to this problem, we introduce the Softmax for Arbitrary Label Trees (SALT) framework, an approach that employs a single, robust model to efficiently manage a broad spectrum of labels. The SALT framework harnesses hierarchical relationships to segment a vast range of anatomical landmarks, reflecting the natural tree-like organization of the human anatomy. By employing conditional probabilities, this framework models the intricate relationships between these landmarks, capturing the complex network of connections among various anatomical structures. In this study, an application of the SALT framework to 3-dimensional CT scans using a nnUNet\citecmd{isensee_nnu-net_2021} architecture is presented. However, the flexibility of this framework allows for its adaptation to other contexts, accommodating both 2D and 3D imaging across diverse image types, and could be used with a wide range of deep learning algorithms. This adaptability underscores the SALT framework's potential as a universal tool for medical image analysis. This optimization aims to resolve existing bottlenecks and substantially improve the utility of CT scan data in real-time clinical settings, facilitating the seamless operation of the segmentation algorithm within the hospital's data acquisition process.

\section{Materials and Methods}
\label{materials-and-methods}

\subsection{Datasets}
\label{datasets}

\begin{figure}
\includegraphics[width=\textwidth]{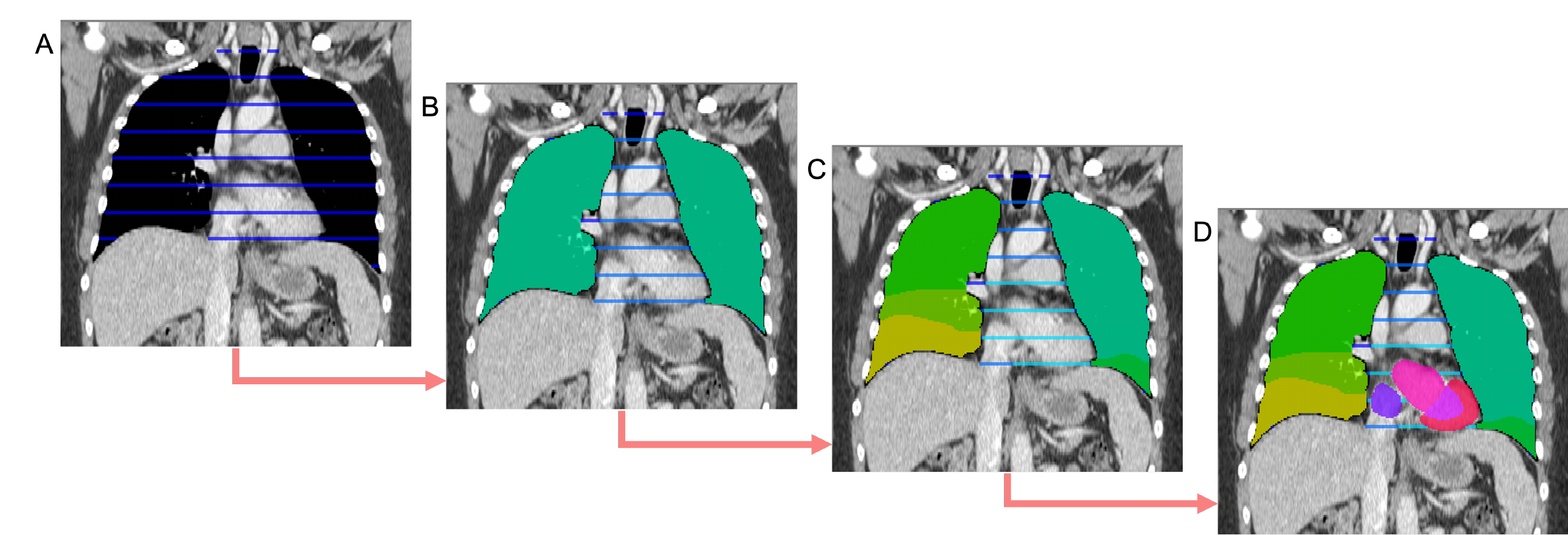}
\captionwithtitle{Example of a hierarchical segmentation}{A) In the thorax, the biggest region is the thoracic cavity. B) The thoracic cavity can then be subdivided into lungs and mediastinum. C) The mediastinum has a further subregion the pericardium, and the lungs can be subdivided into lower/middle/upper left/right lobes. D) The pericardium contains the heart, which can be subdivided into myocardium and left/right ventricle/atrium. The full segmentations originate from the TotalSegmentator, while the sparse segmentations are from SAROS.}
\label{fig:hierarchical}
\end{figure}

This study used a selection of datasets available on The Cancer Imaging Archive (TCIA)\citecmd{clark_cancer_2013} to train and evaluate the SALT approach. For the training, 750 CT scans from the Sparsely Annotated Region and Organ Segmentation (SAROS)\citecmd{clark_cancer_2013,koitka_saros_2024} dataset were used (600 for training and 150 for validation). In this dataset, the segmentations target anatomical landmarks that are relevant for body composition analysis (BCA)\citecmd{haubold_boa_2024,koitka_fully_2021}. The annotations cover a wide range of areas such as the abdominal and thoracic cavities, bones, brain, mediastinum, muscles, pericardium, spinal cord, and subcutaneous tissue. In addition to these annotations, segmentations of organs, vessels, and specific muscles and bones were generated using Version 1 of the TotalSegmentator models\citecmd{wasserthal_totalsegmentator_2023, isensee_nnu-net_2021,wasserthal_dataset_2022} for the same dataset of 750 scans. The TotalSegmentator predictions were then fused with the SAROS annotations to create a single dataset of 750 scans containing all labels. For SAROS, smaller labels such as the thyroid, submandibular, and parotid glands were not included in the final segmentation. The SAROS segmentations encompass larger and more general areas, so the TotalSegmentator predictions were superimposed on the SAROS labels, as a subclassification of larger areas. This fusion is illustrated in \Cref{fig:hierarchical}, which highlights the natural tree-like organization of the human body. For example, the body encompasses the thoracic cavity, which itself includes organs like the lungs and heart. These organs, in turn, can be subdivided further into more specific segments, such as the lobes of the lungs and the atria and ventricles of the heart.

For the evaluation, an additional independent test set of 150 CT scans from SAROS and other publicly available datasets were used: CT Volumes with Multiple Organ Segmentations (CT-ORG)\citecmd{clark_cancer_2013,rister_ct-org_2020,bilic_liver_2023} Fast and Low-resource Semi-supervised Abdominal Organ Segmentation (FLARE22)\citecmd{ma_unleashing_2023,ma_miccai_2023}, Lung CT Segmentation Challenge (LCTSC)\citecmd{clark_cancer_2013,yang_data_2017,yang_autosegmentation_2018}, Lung Nodule Analysis 2016 (LUNA16)\citecmd{armato_iii_lung_2011,tang_automatic_2019}, and Whole Abdominal Organ Dataset (WORD)\citecmd{luo_word_2022}. However, the \qquote{gallbladder} label from the WORD dataset was omitted because the masks were either very few points and not an actual segmentation, a segmentation of the common bile duct, or they were segmentations of a tumor (see \Cref{fig:gallbladder} of \Cref{add-figures}). The final composition of the evaluation dataset, together with the number of cases and the used labels, is reported in \Cref{tab:datasets}.

\begin{table}[!b]
\captionwithtitle{Datasets used for the evaluation}{Multiple datasets from TCIA were used for evaluating SALT:  CT Volumes with Multiple Organ Segmentations (CT-ORG), Fast and Low-resource Semi-supervised Abdominal Organ Segmentation (FLARE22), Lung CT Segmentation Challenge (LCTSC), Lung Nodule Analysis 2016 (LUNA16),  Sparsely Annotated Region and Organ Segmentation (SAROS), and Whole Abdominal Organ Dataset (WORD). For SAROS, an independent test set was used for the evaluation. For each dataset, the number of included CT scans and the labels used for evaluation are reported.}
\begin{tabularx}{\linewidth}{| l | l | X |}
\hline
\textbf{Dataset} & \textbf{Number of Scans} & \textbf{Labels}\\
\hline
CT-ORG\citecmd{clark_cancer_2013,rister_ct-org_2020,bilic_liver_2023} & 21 & Liver, bladder, lungs, kidneys, bone, brain.\\
\hline
FLARE22\citecmd{ma_unleashing_2023,ma_miccai_2023} & 50 & Liver, right kidney, spleen, pancreas, aorta, inferior vena cava, right adrenal gland, left adrenal gland, gallbladder, stomach, duodenum, left kidney.\\
\hline
LCTSC\citecmd{clark_cancer_2013,yang_data_2017,yang_autosegmentation_2018} & 60 & Spinal cord, lung right, lung left, heart.\\
\hline
LUNA16\citecmd{armato_iii_lung_2011,tang_automatic_2019} & 51 & Upper right lobe, middle right lobe, lower right lobe, upper left lobe, lower left lobe.\\
\hline
SAROS\citecmd{clark_cancer_2013,koitka_saros_2023,koitka_saros_2024} & 150 & Subcutaneous tissue, muscles, abdominal cavity, thoracic cavity, bones, mediastinum, pericardium, brain, spinal cord.\\
\hline
WORD\citecmd{luo_word_2022} & 140 & Liver, spleen, kidney left, kidney right, stomach, pancreas, duodenum, colon, intestine, adrenal glands, bladder.\\
\hline
\end{tabularx}
\label{tab:datasets}
\end{table}

\subsubsection{Dataset Postprocessing}
\label{dataset-postprocessing}

The labels from all datasets underwent postprocessing to ensure the hierarchical structure by either merging or splitting the original segmentations. An example of merging is the lung label, which was not present in the TotalSegmentator labels but could be inferred by merging the upper, lower, middle left, and right lobes.
In some cases, splitting was necessary to ensure the tree structure, as each label can only have one parent. For instance, the aorta passes through the mediastinum, pericardium, and abdominal cavity, which would imply three parents. To preserve the tree structure, the aorta label was split into three separate regions: \qquote{aorta thoracica pass pericardium}, \qquote{aorta thoracica pass mediastinum}, and  \qquote{aorta abdominalis}. Similar splitting was performed for the inferior vena cava and pulmonary artery.
Furthermore, in cases where the child labels were insufficient to fully annotate the volume of the parent label, additional labels were created. For instance, the thoracic cavity has two child labels, namely the lungs and the mediastinum (as depicted in \Cref{fig:hierarchical}). However, these two volumes alone do not encompass the entire thoracic cavity. Therefore, an additional label (called \qquote{other}) was introduced to incorporate the remaining voxels and ensure a correct learning process for the model. After postprocessing, there were a total of 145 labels (excluding the root) and 113 leaf nodes. The hierarchical structure of the labels can be viewed in \Cref{fig:landmarks}. 

\begin{figure}
\includegraphics[width=\textwidth]{figures/Figure2.pdf}
\captionwithtitle{Hierarchical labeling for the segmentation of
anatomical landmarks}{The blue labels were generated with the TotalSegmentator, while the pink labels come from the SAROS dataset. The gray labels were also generated: \qquote{body} is the sum of all annotated voxels, while \qquote{background} is all non-annotated voxels. The \qquote{other} classes were created as the parts of the parents which were not annotated. The light pink labels were generated by splitting an existing TotalSegmentator label. Some vertebrae and ribs have been removed from this visualization for a better overview. For some bones and muscles with only left and right children, the two nodes were fused for better visualization.}
\label{fig:landmarks}
\end{figure}

\subsection{Modeling Conditional Probabilities for Trees}
\label{modeling-conditional-probabilities-for-trees}

As visible from \Cref{fig:landmarks}, the hierarchical labels can be represented as an unweighted tree. An arbitrary tree can be defined by its adjacency matrix $A$ of size $N \times N$, where $N$ is the number of nodes and each value $a_{i,j}$ of the matrix represents an edge between parent node $i$ and child node $j$. Since the tree is unweighted, $A$ has only binary elements\citecmd{busato_graph_2017}, and  $a_{ij} = 1$ represents a connection between parent node $i$ and child node $j$.

\begin{figure}
\includegraphics[width=\textwidth]{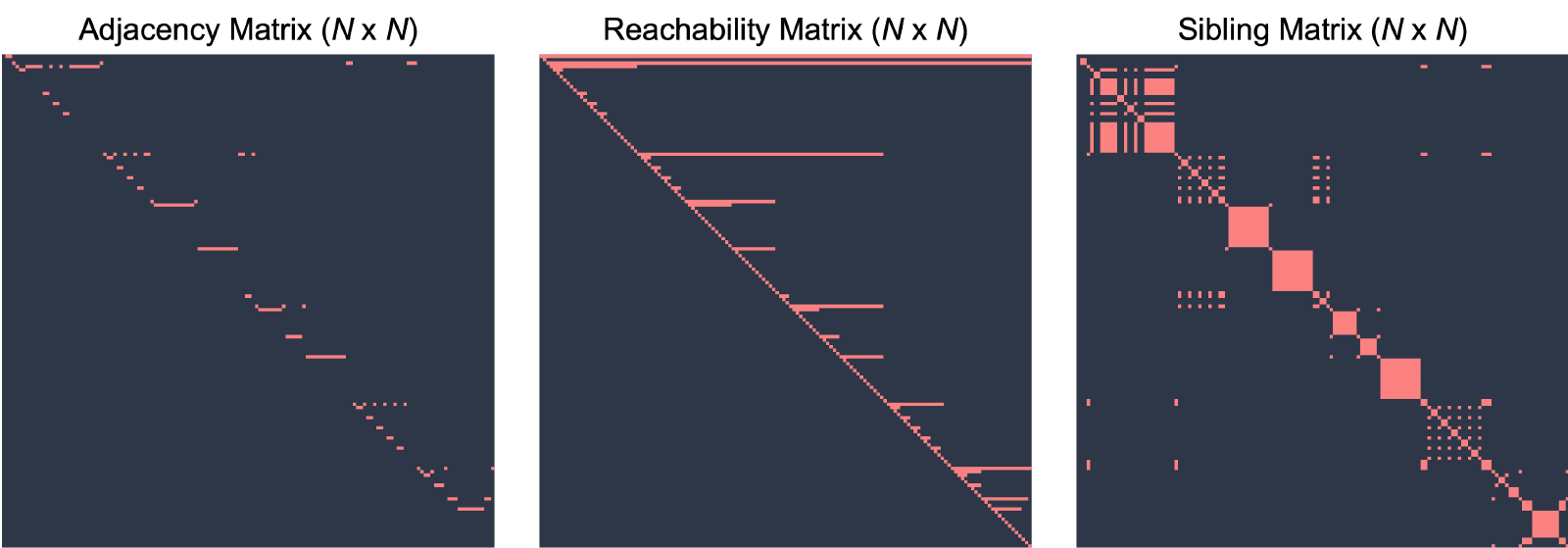}
\captionwithtitle{Different representations of the tree class
hierarchy}{From left to right: (1) The adjacency matrix encodes directed edges from parent to child nodes. (2) The reachability matrix encodes all nodes in the path from the root node to a specific node. (3) The sibling matrix encodes all sibling nodes for a specific node.}
\label{fig:matrices}
\end{figure}

Another useful structure is the reachability matrix $R$, which encodes in row $i$ all the nodes $j$ that can be reached from node $i$, and in column $j$ all the traversed nodes between the root and node $j$. This matrix is also binary and it can be derived from the adjacency matrix using matrix multiplications and additions:
\begin{alignat*}{2}
R = \sum^H_{v=0} A^v = A^0 + A^1 + \ldots + A^H,
\end{alignat*}
where $H$ is the height of the arbitrary tree. Each power $A^v$ represents the nodes that can be reached from any node with a path of size $v$, so the matrix $A^{H + 1}$ will be a zero matrix, as no two nodes have a distance that is larger than $H$.
Another component is the sibling matrix $S$, which is also binary and encodes all local neighbors on the same level:
\begin{alignat*}{2}
S = A^T \times A
\end{alignat*}
An example for all three matrices, $A$, $R$, and $S$, based on the employed dataset and the associated label structure is shown in \Cref{fig:matrices}.

Using these matrices, the assignment of a voxel of the body to a specific class can be represented through conditional probabilities based on the tree structure. For instance, the probability of a voxel to belong to the right lower lobe of the lung depends on the probabilities of it belonging to the right lung, the lungs as a whole, the thoracic cavity, and the body. This concept is analogous to the one of Bayesian networks\citecmd{russell_probabilistic_2010}, which are acyclic directed graphs where each node contains probability information, and edges represent a direct influence between the nodes. In a Bayesian network, all relationships are expressed using conditional probabilities and can be simplified using the chain rule\citecmd{russell_probabilistic_2010}. This can also be applied to our hierarchical tree, but a normalization step needs to be added to ensure that each node represents a probability. This can be done using a softmax function\citecmd{bridle_probabilistic_1990}, which is commonly employed as an activation function in neural networks for multi-class classification tasks. Its outputs are probabilities that indicate the likelihood of the input being assigned to each class, and its sum is 1, just like for probabilities.

These concepts can be used to build a deep learning model that uses conditional probabilities and the softmax function as activation layer for segmentation. The model takes an input and outputs a feature map $x$ with $N$ channels, which is the number of nodes of the hierarchical tree. To enforce the hierarchical relationships, a final probability function for each class $c$ can serve as the activation layer.

The probability $P(y = c | x)$ that the final class $y$ corresponds to class $c$ can be computed using the chain rule as the product of the probabilities from the root node to node 40. The probability of each node is normalized using the softmax for each sibling group, obtaining the following formula:
\begin{alignat*}{2}
P(y=c|x) &= \prod^{R_c}_{i=1} P(y=i|x \bigcap^{i-1}_{j=1} y = j)\\
&=\prod_{i \in R_c} \frac{e^{-x_i}}{\sum_{k \in S_i} e^{-x_k}},
\end{alignat*}
Here, $R_c$ denotes the column $c$ of $R$, which contains the nodes in the path between the root node and node $c$. $S_i$ represents the siblings of node $i$, and $x_i$ and $x_j$ respectively correspond to the feature maps associated with node $i$ and $j$. The resulting probabilities of all leaf nodes sum up to one, analogously to the softmax function. 
To illustrate an example of the probability chain, consider the process of determining the likelihood that a voxel is part of the left hip bone. This requires a sequential evaluation of the voxel's probability of being within the body, within the bone, then associated with hip bones, and finally identified as part of the specific left hip bone, which can be expressed as:
\begin{alignat*}{2}
P(y=\text{hip left}|x) &= P(y=\text{body}) \\
&\ \cdot\ P(y=\text{bones}| x \cap y = \text{body}) \\
&\ \cdot\ P(y=\text{hip}| x \cap y = \text{body} \cap y = \text{bones}) \\
&\ \cdot\ P(y=\text{hip left}| x \cap y = \text{body} \cap y = \text{bones} \cap y = \text{hip}).
\end{alignat*}

\subsection{Model Preprocessing and Training}
\label{model-preprocessing-and-training}

\begin{figure}[!ht]
\includegraphics[width=\textwidth]{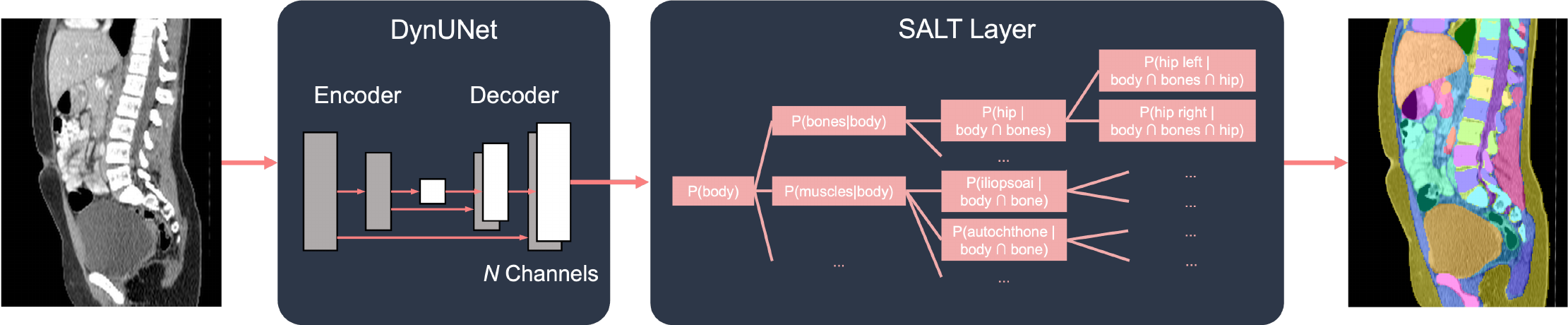}
\captionwithtitle{Overview of the SALT architecture}{A DynUNet model is
used to output feature maps containing $N$ channels, where $N$
is the number of nodes of the tree. The SALT layer builds conditional
probabilities from the nodes of the tree and can be used analogously to
a softmax function to create a segmentation.}
\label{fig:overview}
\end{figure}

The SALT architecture in this paper consists of a DynUNet from the MONAI framework\citecmd{monai_consortium_monai_2022} (version 1.1, PyTorch\citecmd{paszke_pytorch_2019} version 1.14), which is a reimplementation of the architecture utilized by nnUNet\citecmd{isensee_nnu-net_2021}. The output feature map of the DynUNet model has as many channels as there are nodes in the hierarchical tree. Due to SALT’s flexibility, the DynUNet could be substituted by any other model. The previously described probability functions were used to create an activation layer, which was used to generate the final probabilities, as illustrated in \Cref{fig:overview}.

Initially, all CT scans underwent several pre-processing steps: Left-Posterior-Inferior voxel reorientation, resampling to a voxel spacing of 1.5x1.5x1.5 mm, normalization of intensity values within the Hounsfield Units to scale the range from -1024 to 1024 to between 0 and 1,  and random crops of size 192x192x48. The model was trained for 1000 epochs with 600 CT scans for training and 150 for validation. The AdamW optimizer\citecmd{loshchilov_decoupled_2019} was used with an initial learning rate of 0.00025 and a weight decay of 0.00005. The learning rate was reduced during training using a cosine function\citecmd{loshchilov_sgdr_2017}. 

The model used a hybrid loss based on categorical cross-entropy loss\citecmd{murphy_information_2022} and Dice loss\citecmd{milletari_v-net_2016}, which are commonly used for medical segmentation tasks\citecmd{isensee_nnu-net_2021}. In contrast to the common use of these losses, SALT optimizes multiple classes at the same time by constructing an encoding of the nodes using the reachability matrix. Let $R$ be the reachability matrix of size $(N,N)$, where $N$ is the number of nodes in the tree, and let $y$ be the ground truth label with $|V|$ elements (each corresponding to a voxel's label). We construct a new label $y'$ of size $(|V|, N)$ by indexing the reachability matrix $R$ using the ground truth $y$. This operation maps each voxel label to its corresponding encoding column in $R$, or in other words, $y'$ corresponds to the traversed nodes from the root to the voxel's label. Considering the prediction $\hat{y}$ of size $(|V|, N)$, the cross-entropy and the Dice loss can be computed using $y'$ and $\hat{y}$. Unlike the conventional softmax approach, this method optimizes each node along with all its ancestors, which is a result of the label encoding and the implementation of chained conditional probabilities within the activation layers.

For the calculation of the evaluation metrics during training, a similar approach that uses the tree structure to encode each node was used. Technical details about this method are presented in \Cref{implementation-details}. Furthermore, the trained model and the code are available for review on GitHub under the following link: \url{https://github.com/UMEssen/SALT}.

\subsection{Evaluation}
\label{evaluation}

The model's performance was assessed with the Dice score\citecmd{dice_measures_1945} and with the normalized surface Dice score (NSD)\citecmd{nikolov_clinically_2021}. The NSD measures the frequency at which the surface distance between volumes is under 3 mm, a metric previously employed by TotalSegmentator\citecmd{wasserthal_totalsegmentator_2023}. For a better comparison with existing models, the same metrics were also utilized to evaluate the predictions from Version 2 of the TotalSegmentator\citecmd{wasserthal_totalsegmentator_2023,isensee_nnu-net_2021} on the datasets from \Cref{tab:datasets}. Additionally, 95\% confidence intervals (reported in square brackets) were computed by bootstrapping the scores of the evaluated CT scans. The bootstrapping was performed with 1000 iterations and the 2.5 and the 97.5 percentiles were utilized as upper and lower bound of the confidence intervals. To evaluate the variations in scores across datasets, the distributions of the Dice and NSD scores were compared for datasets that share the same organ labels.

For a better comparison, the evaluation datasets were also used to generate predictions from Version 2 of the TotalSegmentator\citecmd{wasserthal_totalsegmentator_2023,isensee_nnu-net_2021} and the same evaluation scores were computed. Moreover, the speed of the model was evaluated in comparison with Version 1 and Version 2 of the TotalSegmentator and across multiple speed settings. For TotalSegmentator Version 1, tests were carried out on models with isotropic spacings of 1.5mm and 3mm. In the case of Version 2, models with isotropic spacings of 1.5mm, 3mm, and 6mm were examined.  All experiments were performed by running the models on a single NVIDIA RTX A6000 card\citecmd{nvidia_corporation_nvidia_2024}. 

\section{Results}
\label{results}

\subsection{Segmentation Evaluation}
\label{segmentation-evaluation}

The trained model showed a Dice of $0.891 [0.887, 0.896]$ and an NSD of $0.931 [0.927, 0.936]$. An overall score for the different datasets is presented in \Cref{tab:summary} together with the scores obtained by Version 2 of the TotalSegmentator. The Dice scores and the NSD scores for the datasets are reported in \Cref{tab:ctorg}, \Cref{tab:flare22}, \Cref{tab:lctsc}, \Cref{tab:luna16}, \Cref{tab:saros} and \Cref{tab:word} of \Cref{add-tables}. Notably, the lungs, the liver, the spleen, and the stomach achieved the best scores across the different datasets. 

\begin{table}[!b]
\captionwithtitle{Summary results of the model's performance for the
datasets}{The evaluation is performed in terms of Dice and Normalized Surface Dice (NSD) for both SALT and Version 2 of the TotalSegmentator (TSV2). The 95\% confidence intervals were computed using 1000 rounds of bootstrapping, and the interval is reported in square brackets.}
{\footnotesize
\begin{tabularx}{\linewidth}{| Y | Y | Y | Y | Y | Y | Y | Y |}
\hline
\textbf{Metric} & \textbf{CT-ORG} & \textbf{FLARE22} & \textbf{LCTSC} & \textbf{LUNA16} & \textbf{SAROS} & \textbf{WORD}\\
\hline
SALT (Dice) & 0.891 \par
[0.869, 0.906] & 0.849 \par
[0.844, 0.854] & 0.908 \par
[0.902, 0.914] & 0.93 \par
[0.919, 0.938] & 0.929 \par
[0.924, 0.933] & 0.844 \par
[0.839, 0.85]\\
\hline
TSV2 (Dice) & 0.917 \par
[0.898, 0.931] & 0.892 \par
[0.888, 0.897] & 0.937 \par
[0.931, 0.942] & 0.952 \par
[0.943, 0.959] & 0.852 \par
[0.846, 0.857] & 0.84 \par
[0.834, 0.846]\\
\hline
SALT (NSD) & 0.884 \par
[0.863, 0.905] & 0.947 \par
[0.942, 0.952] & 0.886 \par
[0.872, 0.899] & 0.908 \par
[0.89, 0.92] & 0.98 \par
[0.975, 0.984] & 0.909 \par
[0.903, 0.915]\\
\hline
TSV2 (NSD) & 0.916 \par
[0.895, 0.933] & 0.954 \par
[0.949, 0.959] & 0.945 \par
[0.935, 0.954] & 0.934 \par
[0.925, 0.943] & 0.955 \par
[0.949, 0.959] & 0.906 \par
[0.899, 0.912]\\
\hline
\end{tabularx}
}
\label{tab:summary}
\end{table}

\begin{figure}
\includegraphics[width=\textwidth]{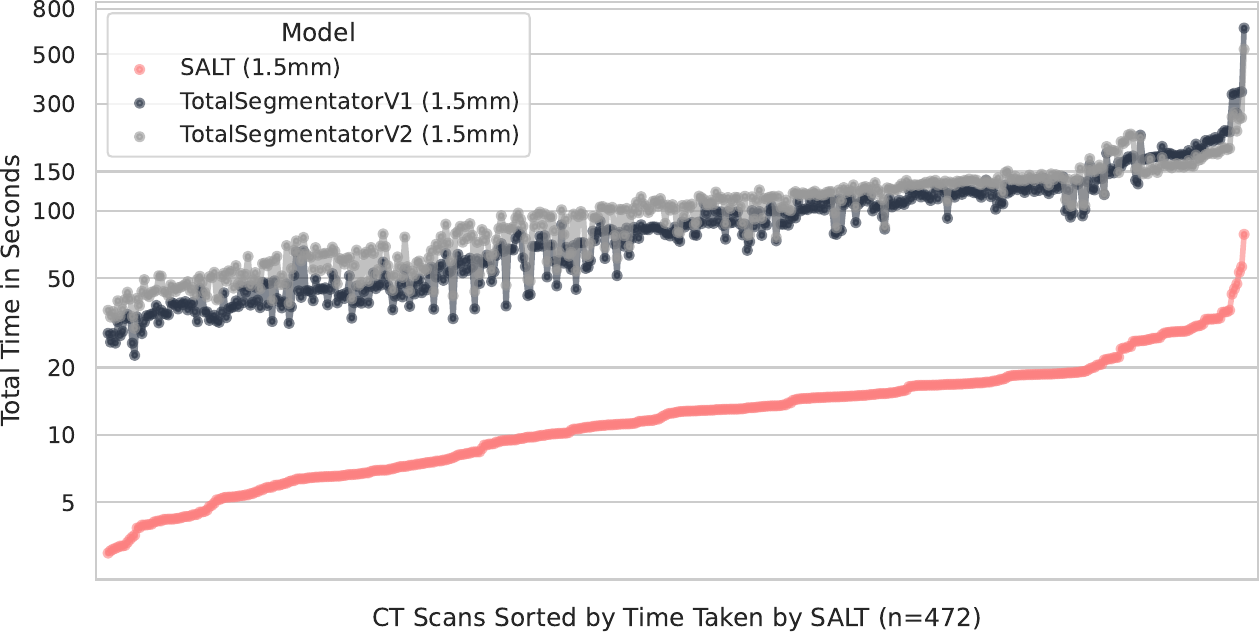}
\captionwithtitle{Comparison between the speed of SALT and the TotalSegmentator}{Version 1 and 2 of the TotalSegmentator were run on the same set of 472 CT scans from \Cref{tab:datasets}. The TotalSegmentator models and SALT were trained on 1.5mm isotropic spacing.}
\label{fig:speed}
\end{figure}

Additionally, an evaluation of the speed of the model at inference time was also performed, which can be reviewed in \Cref{tab:time}. It is relevant to make a distinction between the inference time (the time the model takes to make a prediction) and the total time, as a large portion of the inference time is just spent postprocessing and storing the result. In \Cref{fig:speed}, a comparative analysis of SALT's speed against both Version 1 and Version 2 of the TotalSegmentator is presented. This comparison clearly demonstrates that SALT consistently outperforms its counterparts, showing a notable speed advantage, especially with larger CT scans. Additionally, a speed comparison was conducted between SALT and the faster variants of TotalSegmentator that utilize lower spacing, with the findings detailed in \Cref{fig:speed2} of \Cref{add-figures}. These \qquote{fast} alternatives of TotalSegmentator, which use a single model instead of five, also exhibit better speed. In these comparisons, the 3mm model from Version 2 of TotalSegmentator was consistently slower than or comparable to SALT. However, for larger images, the 3mm model of Version 1 and the 6mm model of Version 2 were faster.

\begin{figure}
\includegraphics[width=\textwidth]{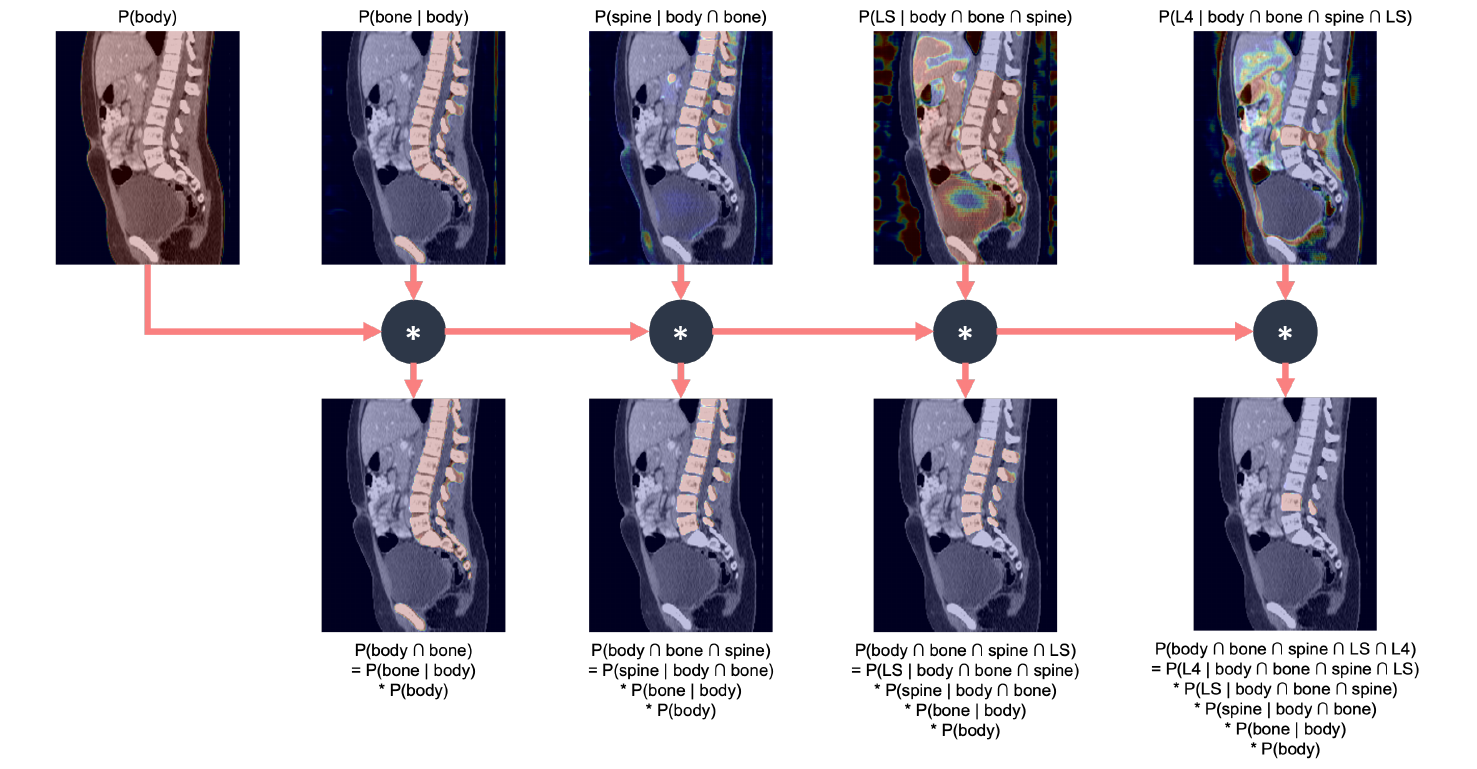}
\captionwithtitle{Visualization of conditional probability maps for segmenting the L4 vertebra using the trained model}{In the visualization, it is showcased that the model does not directly predict the target class, but uses conditional probabilities to create the predictions. LS: Lumbar Spine.}
\label{fig:conditional}
\end{figure}

Furthermore, the inner workings of the model can be better understood by examining the conditional probabilities used at various stages. As an example, the process of predicting the vertebra L4 is illustrated in \Cref{fig:conditional}. This visualization effectively demonstrates how the model relies on these conditional probabilities to formulate its predictions.

\subsection{Failure Analysis}
\label{failure-analysis}

In many cases, the results for the same classes varied across the datasets. To investigate these discrepancies, statistical tests were performed on datasets sharing the same labels to discover whether there were significant differences in the results. The results are reported in \Cref{tab:statistics} of \Cref{add-tables}. 

\begin{figure}
\includegraphics[width=\textwidth]{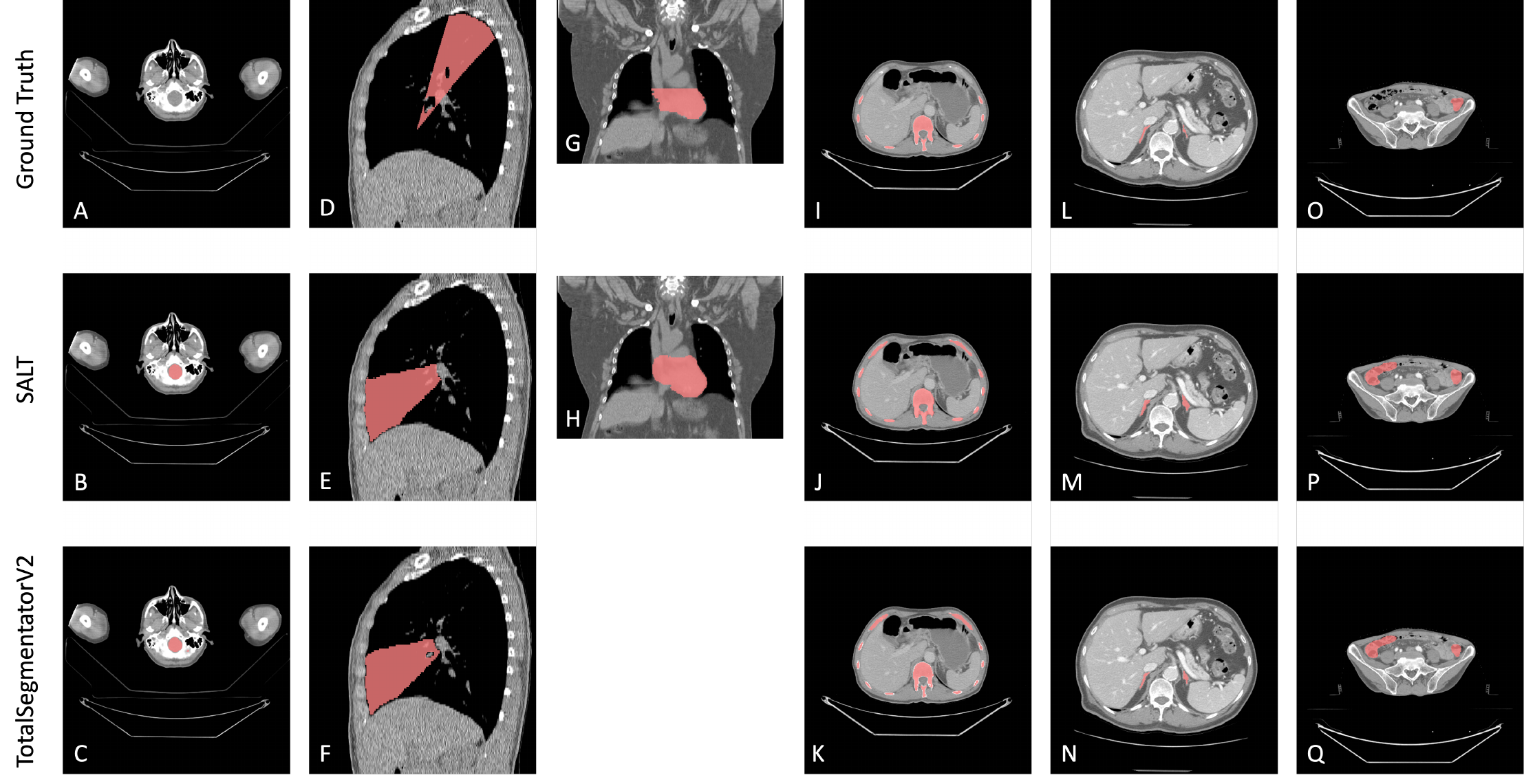}
\captionwithtitle{Comparison between the ground truth (first row), SALT (second row) and Version 2 of the TotalSegmentator (third row)}{The comparisons across various classes and datasets are systematically organized as follows: Figures A, B, and C focus on the comparison of the brain class from the CT-ORG dataset. Figures D, E, and F examine the middle lobe of the LUNA16 dataset. In Figures G and H, the analysis shifts to the pericardium from the LCTSC dataset. The bone class from the CT-ORG dataset is compared in Figures I, J, and K. Figures L, M, and N represent the adrenal glands from the FLARE22 datasets. Finally, Figures O, P, and Q compare the colon class from the WORD dataset.}
\label{fig:comparison}
\end{figure}

An example of this discrepancy is the brain class, which however did not present a significant difference in \Cref{tab:statistics} due to the few CT scans with this label. For the SAROS dataset, the brain class achieved a Dice score of 0.758, while for the CT-ORG dataset, it only obtained a Dice score of 0.486. This is because two CT scans of the CT-ORG dataset were classified as containing a brain segmentation by the model, but only one was annotated as such in CT-ORG. The existing segmentation achieved a Dice score of 0.973. The other case should also have been segmented, as the image, despite not showing the entire brain, still includes a visible brainstem (\Cref{fig:comparison}B). The TotalSegmentator also produced a similar segmentation as our model (\Cref{fig:comparison}C). In SAROS, in nine cases where the brain was not part of the CT scan, the model incorrectly predicted the brain class in other areas of the body. 

In the LUNA16 dataset, the lung segmentations consistently achieved good results. However, for the middle lobe, there were some significant differences in the segmentations. In fact, sometimes the middle lobe was wrongly segmented, and in \Cref{fig:comparison}E and \Cref{fig:comparison}F it is possible to see that for these examples, the TotalSegmentator produced similar segmentations as our model. This is also further proven by the 0.848 Dice score of the TotalSegmentator for this class (\Cref{tab:luna16} of \Cref{add-tables}).

For the pericardium class, the annotations from the LCTSC datasets followed another definition and resulted in an overall Dice score of 0.894 compared to a Dice score of 0.952 for the SAROS dataset. According to the LCTSC guidelines, the heart is contoured around the pericardial sack. This annotation is thus not compatible with the ones produced by our model, which segments the entire area within the pericardial sack (\Cref{fig:comparison}H). 

For the bone class, where the model achieved a Dice score of 0.911 for the SAROS dataset, the Dice score for the CT-ORG dataset was 0.872. Visually, it can be seen that the model tends to include more contours than the ones from this dataset (\Cref{fig:comparison}I, \Cref{fig:comparison}J, and \Cref{fig:comparison}K). In particular, our segmentations also include the cortical bone, while the TotalSegmentator and the CT-ORG segmentations tend to leave it out. Moreover, in the SAROS dataset, cartilage was annotated as bone in the rib cage, and since the rib cage is considered part of the bones, this causes a discrepancy in the segmentations. In Version 1 of the TotalSegmentator, the segmentation of the cartilage was not present, but it was added in Version 2 of the tool, as can be seen in \Cref{fig:comparison}K.

The adrenal glands class consistently achieved the lowest Dice scores between 0.65 and 0.706. Visually, our model tends to segment a larger area that also contains the organ’s wall (\Cref{fig:comparison}L, \Cref{fig:comparison}M, and \Cref{fig:comparison}N). This is also proven by the better NSD scores (between 0.887 and 0.957). For the TotalSegmentator, this class also produced similar results in the WORD dataset but had a better Dice score for the FLARE22 dataset (\Cref{tab:flare22} of \Cref{add-tables}).

For the colon segmentation, the model occasionally encountered challenges in differentiating between the colon and the small bowel. This issue is exemplified in \Cref{fig:comparison}O and \Cref{fig:comparison}P, where the patient underwent a right hemicolectomy, resulting in the removal of the ascending colon and a portion of the transverse colon. Due to the presence of significant air within the small bowel, the model erroneously identified it as part of the colon. However, as can be seen in \Cref{fig:comparison}Q, the TotalSegmentator also made this mistake, even resulting in a lower Dice score for this class compared to SALT (0.773 vs. 0.808).

\section{Discussion}
\label{discussion}

In this work, we propose the SALT framework to provide a single activation layer for hierarchical probability modeling. While we have illustrated its application in medical imaging with a nnUNet model, the activation layer could be adapted to various hierarchical contexts and could be used with any model. The power of this framework lies in its ability to train a single segmentation model to recognize over 100 labels hierarchically. Throughout the training process, the loss function is computed over the full tree, allowing for the optimization of each node at every stage of training. This hierarchical training ensures that the model respects the natural structure of the data, for instance, ensuring that the colon is identified within the abdominal cavity. Additionally, the proposed model takes an average of 35 ms per slice, meaning that a full whole-body CT scan with 1000 slices can be computed in 35 seconds, which is faster than existing models\citecmd{wasserthal_totalsegmentator_2023}. 

\subsection{Performance Comparison}
\label{performance-comparison}

In general, SALT obtained consistently good results for the kidneys (Dice 0.92, except for the CT-ORG dataset), the liver (Dice above 0.95), the spleen (Dice above 0.92), the stomach (Dice above 0.9), and the lungs (Dice above 0.91, except for the right middle lobe). The most difficult class to predict in terms of Dice was the adrenal glands (Dice scores of 0.65-0.70), which however showed good results for the NSD (0.88-0.95). The segmentation produced by the model tried to take more contours into account than the external dataset, yielding lower Dice scores for classes that do not account for many voxels (such as the adrenal glands). 

On the SAROS dataset, SALT achieved similar results to a nnUNet-based network for BCA trained on similar classes\citecmd{haubold_boa_2024}. Specifically, SALT achieved a better Dice score for the mediastinum class (0.95 vs. 0.84) and maintained a performance difference of no more than 0.05 decimal points compared to the BCA nnUNet network across most classes. The only exception was the brain class, where nnUNet achieved a higher score of 0.97 compared to SALT's 0.75. Upon comparing SALT's results with those of TotalSegmentator\citecmd{wasserthal_totalsegmentator_2023} for the other datasets, it was observed most classes had a difference of at most 0.05. The only exceptions were the adrenal glands and the vena cava inferior in the FLARE22 dataset, the kidneys and the urinary bladder in the CT-ORG dataset, and the brain in the SAROS dataset. Additionally, larger errors were reported in classes belonging to the gastrointestinal tract, a region traditionally challenging to segment accurately\citecmd{fu_novel_2018,wang_bowelnet_2023,siddiqui_differentiating_2019}. In fact, annotating structures such as the small and large intestines and the duodenum, presents a challenge due to their inherent complexity. These anatomical regions are notorious for their considerable variability, lack of consistent positioning, and occasional absence of clearly delineated boundaries\citecmd{fu_novel_2018,wang_bowelnet_2023}. 

Moreover, the scores presented significant differences across the datasets, exemplifying the heterogeneity of annotation guidelines and the inherent variability of human anatomy. This is illustrated in \Cref{fig:comparison} and showcased by the statistical comparisons of \Cref{tab:statistics} of \Cref{add-tables} and by the results of the TotalSegmentator, with a Dice score of 0.835 for adrenal glands in the FLARE22 dataset and a score of 0.624 in the WORD dataset.

\subsection{Clinical Applicability}
\label{clinical-applicability}

The extraction of biomarkers from medical segmentations is an active field of research that has a variety of clinical applications, such as making accurate diagnoses\citecmd{aromiwura_artificial_2023}, monitoring a patient's well-being\citecmd{jung_association_2023,alderuccio_quantitative_2023}, or predicting overall survival\citecmd{ito_spleen_2023,khoshpouri_quantitative_2019}. This also includes BCA biomarkers, which are also relevant for disease progression\citecmd{jung_association_2023,mason_respiratory_2021,ko_change_2022} and predicting patient survival outcomes\citecmd{hosch_biomarkers_2022,keyl_deep_2023}. Moreover, segmentations can also be used to predict contrast phases\citecmd{baldini_addressing_2023}. Given the many applications and the high volume of CT scans generated daily, models for segmentation and biomarker extraction need to prioritize speed and efficiency. SALT's fast processing suggests that, compared to other models, it could enable quicker biomarker extraction, resulting in increased throughput. This speed could facilitate its integration into clinical workflows, enabling the automatic calculation of essential biomarkers each time a patient undergoes a CT scan. 

\subsection{Limitations \& Future Work}
\label{limitations-future-work}

A limitation of this study is the tendency of the model to include the walls of organs and vessels in the annotations, which differs from the annotations of the other datasets. While this approach may be advantageous for certain classes, such as bones, where it includes the cortical bone area, it may be disadvantageous for others, such as adrenal glands, where it may inadvertently include abdominal fat. This may be due to differences in annotation between the large-scale segmentations of the SALT dataset (which typically include the walls) and the TotalSegmentator predictions. Moreover, SALT also performed worse in the aorta and inferior vena cava classes (0.89 and 0.859 respectively) compared to TotalSegmentator (0.936 and 0.912 respectively). This discrepancy might be attributed to the necessity of dividing these structures into multiple regions to ensure the tree structure, which may have confused the model. This limitation arises from the required tree-like class hierarchy of the model, as not all classes are strictly assignable to a single parent node. A graph-like topology, represented by a directed acyclic graph, would greatly extend the representation capabilities, as a node could potentially have multiple parent nodes, which could be implemented as a union of all parent probabilities. However, the overall softmax properties would no longer hold, as the probabilities would not sum up to one. Another interesting extension would be to define the hierarchical tree using existing ontologies, such as the Foundational Model of Anatomy\citecmd{rosse_foundational_2008} or SNOMED\citecmd{el-sappagh_snomed_2018}, to provide standardization.

An additional extension could be used for multi-label segmentation, where each voxel gets more than one class assigned. This is especially important for overlapping concepts, such as that subcutaneous tissue or muscle can be present in all parts of the body (arms, head, legs, and torso). Moreover, this work was also intended to work with multiple datasets, such that, without any modifications, a model could be trained using different sources without needing to fuse different datasets on one single segmentation. 

Furthermore, future work should aim to refine the segmentation capabilities of the model to ensure that organ walls are only included in the annotations when relevant, which raises the larger problem of unifying datasets annotated with different annotation guidelines. The introduction of advanced post-processing methods could also improve the quality of the resulting segmentations, leading to more accurate and clinically useful outcomes.

\section{Conclusion}
\label{conclusion}

In conclusion, the presented SALT framework offers a new approach to medical imaging by utilizing the hierarchical nature of the human anatomy to achieve comprehensive and efficient segmentations across 113 body regions. The model's capability to process a whole-body CT scan in an average of 35 seconds paves the way for integration into clinical workflows, enabling the immediate computation of crucial biomarkers upon performing a CT scan. While this work primarily targeted the segmentation of body regions in CT scans, SALT defines an activation function that can be used for any application or domain with a hierarchical structure. Future enhancements will focus on refining the segmentation accuracy, exploring graph-like topologies for more complex anatomical structures, and extending the model's application to multi-label segmentation and multi-dataset integration.

\printbibliography

@misc{loshchilov_decoupled_2019,
	title = {Decoupled Weight Decay Regularization},
	url = {http://arxiv.org/abs/1711.05101},
	doi = {10.48550/arXiv.1711.05101},
	number = {{arXiv}:1711.05101},
	publisher = {{arXiv}},
	author = {Loshchilov, Ilya and Hutter, Frank},
	urldate = {2023-06-13},
	date = {2019-01-04},
	eprinttype = {arxiv},
	eprint = {1711.05101 [cs, math]},
}

@misc{loshchilov_sgdr_2017,
	title = {{SGDR}: Stochastic Gradient Descent with Warm Restarts},
	url = {http://arxiv.org/abs/1608.03983},
	doi = {10.48550/arXiv.1608.03983},
	shorttitle = {{SGDR}},
	number = {{arXiv}:1608.03983},
	publisher = {{arXiv}},
	author = {Loshchilov, Ilya and Hutter, Frank},
	urldate = {2023-06-12},
	date = {2017-05-03},
	eprinttype = {arxiv},
	eprint = {1608.03983 [cs, math]},
}

@inproceedings{milletari_v-net_2016,
	title = {V-Net: Fully Convolutional Neural Networks for Volumetric Medical Image Segmentation},
	doi = {10.1109/3DV.2016.79},
	shorttitle = {V-Net},
	eventtitle = {2016 Fourth International Conference on 3D Vision (3DV)},
	pages = {565--571},
	booktitle = {2016 Fourth International Conference on 3D Vision (3DV)},
	author = {Milletari, Fausto and Navab, Nassir and Ahmadi, Seyed-Ahmad},
	date = {2016-10},
}

@incollection{murphy_information_2022,
	title = {Information Theory},
	isbn = {978-0-262-04682-4},
	url = {probml.ai},
	pages = {203--205},
	booktitle = {Probabilistic Machine Learning: An introduction},
	publisher = {{MIT} Press},
	author = {Murphy, Kevin P.},
	date = {2022},
}

@incollection{russell_probabilistic_2010,
	location = {Upper Saddle River},
	edition = {3rd ed},
	title = {Probabilistic Reasoning},
	isbn = {978-0-13-604259-4},
	series = {Prentice Hall series in artificial intelligence},
	pages = {510--515},
	booktitle = {Artificial intelligence: a modern approach},
	publisher = {Prentice Hall},
	author = {Russell, Stuart J. and Norvig, Peter},
	date = {2010},
	langid = {english},
}

@inproceedings{bridle_probabilistic_1990,
	location = {Berlin, Heidelberg},
	title = {Probabilistic Interpretation of Feedforward Classification Network Outputs, with Relationships to Statistical Pattern Recognition},
	isbn = {978-3-642-76153-9},
	doi = {10.1007/978-3-642-76153-9_28},
	series = {{NATO} {ASI} Series},
	pages = {227--236},
	booktitle = {Neurocomputing},
	publisher = {Springer},
	author = {Bridle, John S.},
	editor = {Soulié, Françoise Fogelman and Hérault, Jeanny},
	date = {1990},
	langid = {english},
}

@incollection{busato_graph_2017,
	title = {Graph algorithms on {GPUs}},
	isbn = {978-0-12-803738-6},
	url = {https://linkinghub.elsevier.com/retrieve/pii/B9780128037386000070},
	pages = {163--198},
	booktitle = {Advances in {GPU} Research and Practice},
	publisher = {Elsevier},
	author = {Busato, F. and Bombieri, N.},
	urldate = {2023-06-09},
	date = {2017},
	langid = {english},
	doi = {10.1016/B978-0-12-803738-6.00007-0},
}

@inproceedings{paszke_pytorch_2019,
	title = {{PyTorch}: An Imperative Style, High-Performance Deep Learning Library},
	volume = {32},
	url = {https://proceedings.neurips.cc/paper_files/paper/2019/hash/bdbca288fee7f92f2bfa9f7012727740-Abstract.html},
	shorttitle = {{PyTorch}},
	booktitle = {Advances in Neural Information Processing Systems},
	publisher = {Curran Associates, Inc.},
	author = {Paszke, Adam and Gross, Sam and Massa, Francisco and Lerer, Adam and Bradbury, James and Chanan, Gregory and Killeen, Trevor and Lin, Zeming and Gimelshein, Natalia and Antiga, Luca and Desmaison, Alban and Kopf, Andreas and Yang, Edward and {DeVito}, Zachary and Raison, Martin and Tejani, Alykhan and Chilamkurthy, Sasank and Steiner, Benoit and Fang, Lu and Bai, Junjie and Chintala, Soumith},
	urldate = {2023-06-02},
	date = {2019},
}

@article{keyl_deep_2023,
	title = {Deep learning-based assessment of body composition and liver tumour burden for survival modelling in advanced colorectal cancer},
	volume = {14},
	issn = {2190-6009},
	doi = {10.1002/jcsm.13158},
	pages = {545--552},
	number = {1},
	journaltitle = {Journal of Cachexia, Sarcopenia and Muscle},
	shortjournal = {J Cachexia Sarcopenia Muscle},
	author = {Keyl, Julius and Hosch, René and Berger, Aaron and Ester, Oliver and Greiner, Tobias and Bogner, Simon and Treckmann, Jürgen and Ting, Saskia and Schumacher, Brigitte and Albers, David and Markus, Peter and Wiesweg, Marcel and Forsting, Michael and Nensa, Felix and Schuler, Martin and Kasper, Stefan and Kleesiek, Jens},
	date = {2023-02},
	pmid = {36544260},
	pmcid = {PMC9891942},
}

@article{dice_measures_1945,
	title = {Measures of the Amount of Ecologic Association Between Species},
	volume = {26},
	issn = {1939-9170},
	url = {https://onlinelibrary.wiley.com/doi/abs/10.2307/1932409},
	doi = {10.2307/1932409},
	pages = {297--302},
	number = {3},
	journaltitle = {Ecology},
	author = {Dice, Lee R.},
	urldate = {2023-05-09},
	date = {1945},
	langid = {english},
	note = {\_eprint: https://onlinelibrary.wiley.com/doi/pdf/10.2307/1932409},
}

@misc{wasserthal_dataset_2022,
	title = {Dataset with segmentations of 104 important anatomical structures in 1204 {CT} images},
	url = {https://zenodo.org/record/6802614},
	doi = {10.5281/zenodo.6802614},
	version = {1.0},
	publisher = {Zenodo},
	author = {Wasserthal, Jakob},
	urldate = {2023-05-09},
	date = {2022-07-06},
}

@article{koitka_fully_2022,
	title = {Fully automated preoperative liver volumetry incorporating the anatomical location of the central hepatic vein},
	volume = {12},
	issn = {2045-2322},
	doi = {10.1038/s41598-022-20778-4},
	pages = {16479},
	number = {1},
	journaltitle = {Scientific Reports},
	shortjournal = {Sci Rep},
	author = {Koitka, Sven and Gudlin, Phillip and Theysohn, Jens M. and Oezcelik, Arzu and Hoyer, Dieter P. and Dayangac, Murat and Hosch, René and Haubold, Johannes and Flaschel, Nils and Nensa, Felix and Malamutmann, Eugen},
	date = {2022-10-01},
	pmid = {36183002},
	pmcid = {PMC9526715},
}

@article{koitka_fully_2021,
	title = {Fully automated body composition analysis in routine {CT} imaging using 3D semantic segmentation convolutional neural networks},
	volume = {31},
	issn = {1432-1084},
	doi = {10.1007/s00330-020-07147-3},
	pages = {1795--1804},
	number = {4},
	journaltitle = {European Radiology},
	shortjournal = {Eur Radiol},
	author = {Koitka, Sven and Kroll, Lennard and Malamutmann, Eugen and Oezcelik, Arzu and Nensa, Felix},
	date = {2021-04},
	pmid = {32945971},
	pmcid = {PMC7979624},
}

@article{isensee_nnu-net_2021,
	title = {{nnU}-Net: a self-configuring method for deep learning-based biomedical image segmentation},
	volume = {18},
	rights = {2020 The Author(s), under exclusive licence to Springer Nature America, Inc.},
	issn = {1548-7105},
	url = {https://www.nature.com/articles/s41592-020-01008-z},
	doi = {10.1038/s41592-020-01008-z},
	shorttitle = {{nnU}-Net},
	pages = {203--211},
	number = {2},
	journaltitle = {Nature Methods},
	shortjournal = {Nat Methods},
	author = {Isensee, Fabian and Jaeger, Paul F. and Kohl, Simon A. A. and Petersen, Jens and Maier-Hein, Klaus H.},
	urldate = {2023-07-07},
	date = {2021},
	langid = {english},
}

@article{ito_spleen_2023,
	title = {Spleen volume is a predictor of posthepatectomy liver failure and short-term mortality for hepatocellular carcinoma},
	volume = {408},
	issn = {1435-2451},
	doi = {10.1007/s00423-023-03025-w},
	pages = {297},
	number = {1},
	journaltitle = {Langenbeck's Archives of Surgery},
	shortjournal = {Langenbecks Arch Surg},
	author = {Ito, Takahiro and Tanemura, Akihiro and Kuramitsu, Toru and Murase, Taichi and Kaluba, Benson and Noguchi, Daisuke and Fujii, Tekehiro and Yuge, Takuya and Maeda, Koki and Hayasaki, Aoi and Gyoten, Kazuyuki and Iizawa, Yusuke and Murata, Yasuhiro and Kuriyama, Naohisa and Kishiwada, Masashi and Mizuno, Shugo},
	date = {2023-08-07},
	pmid = {37548783},
}

@article{nowak_fully_2020,
	title = {Fully Automated Segmentation of Connective Tissue Compartments for {CT}-Based Body Composition Analysis: A Deep Learning Approach},
	volume = {55},
	issn = {1536-0210},
	doi = {10.1097/RLI.0000000000000647},
	shorttitle = {Fully Automated Segmentation of Connective Tissue Compartments for {CT}-Based Body Composition Analysis},
	pages = {357--366},
	number = {6},
	journaltitle = {Investigative Radiology},
	shortjournal = {Invest Radiol},
	author = {Nowak, Sebastian and Faron, Anton and Luetkens, Julian A. and Geißler, Helena L. and Praktiknjo, Michael and Block, Wolfgang and Thomas, Daniel and Sprinkart, Alois M.},
	date = {2020-06},
	pmid = {32369318},
}

@article{nowak_end--end_2022,
	title = {End-to-end automated body composition analyses with integrated quality control for opportunistic assessment of sarcopenia in {CT}},
	volume = {32},
	issn = {0938-7994},
	url = {https://www.ncbi.nlm.nih.gov/pmc/articles/PMC9038788/},
	doi = {10.1007/s00330-021-08313-x},
	pages = {3142--3151},
	number = {5},
	journaltitle = {European Radiology},
	shortjournal = {Eur Radiol},
	author = {Nowak, Sebastian and Theis, Maike and Wichtmann, Barbara D. and Faron, Anton and Froelich, Matthias F. and Tollens, Fabian and Geißler, Helena L. and Block, Wolfgang and Luetkens, Julian A. and Attenberger, Ulrike I. and Sprinkart, Alois M.},
	urldate = {2023-08-25},
	date = {2022},
	pmid = {34595539},
	pmcid = {PMC9038788},
}

@article{rister_ct-org_2020,
	title = {{CT}-{ORG}, a new dataset for multiple organ segmentation in computed tomography},
	volume = {7},
	rights = {2020 The Author(s)},
	issn = {2052-4463},
	url = {https://www.nature.com/articles/s41597-020-00715-8},
	doi = {10.1038/s41597-020-00715-8},
	pages = {381},
	number = {1},
	journaltitle = {Scientific Data},
	shortjournal = {Sci Data},
	author = {Rister, Blaine and Yi, Darvin and Shivakumar, Kaushik and Nobashi, Tomomi and Rubin, Daniel L.},
	urldate = {2023-08-25},
	date = {2020-11-11},
	langid = {english},
	note = {Number: 1
Publisher: Nature Publishing Group},
}

@incollection{rosse_foundational_2008,
	location = {London},
	title = {The Foundational Model of Anatomy Ontology},
	isbn = {978-1-84628-885-2},
	url = {https://doi.org/10.1007/978-1-84628-885-2_4},
	pages = {59--117},
	booktitle = {Anatomy Ontologies for Bioinformatics: Principles and Practice},
	publisher = {Springer},
	author = {Rosse, Cornelius and Mejino, José L. V.},
	editor = {Burger, Albert and Davidson, Duncan and Baldock, Richard},
	urldate = {2024-03-27},
	date = {2008},
	langid = {english},
	doi = {10.1007/978-1-84628-885-2_4},
}

@misc{ma_miccai_2023,
	title = {{MICCAI} {FLARE}22 Challenge Dataset (50 Labeled Abdomen {CT} Scans)},
	url = {https://zenodo.org/records/7860267},
	doi = {10.5281/zenodo.7860267},
	version = {1.0},
	publisher = {Zenodo},
	author = {Ma, Jun},
	urldate = {2024-01-03},
	date = {2023-04-24},
}

@article{bilic_liver_2023,
	title = {The Liver Tumor Segmentation Benchmark ({LiTS})},
	volume = {84},
	issn = {13618415},
	url = {http://arxiv.org/abs/1901.04056},
	doi = {10.1016/j.media.2022.102680},
	pages = {102680},
	journaltitle = {Medical Image Analysis},
	shortjournal = {Medical Image Analysis},
	author = {Bilic, Patrick and Christ, Patrick and Li, Hongwei Bran and Vorontsov, Eugene and Ben-Cohen, Avi and Kaissis, Georgios and Szeskin, Adi and Jacobs, Colin and Mamani, Gabriel Efrain Humpire and Chartrand, Gabriel and Lohöfer, Fabian and Holch, Julian Walter and Sommer, Wieland and Hofmann, Felix and Hostettler, Alexandre and Lev-Cohain, Naama and Drozdzal, Michal and Amitai, Michal Marianne and Vivantik, Refael and Sosna, Jacob and Ezhov, Ivan and Sekuboyina, Anjany and Navarro, Fernando and Kofler, Florian and Paetzold, Johannes C. and Shit, Suprosanna and Hu, Xiaobin and Lipková, Jana and Rempfler, Markus and Piraud, Marie and Kirschke, Jan and Wiestler, Benedikt and Zhang, Zhiheng and Hülsemeyer, Christian and Beetz, Marcel and Ettlinger, Florian and Antonelli, Michela and Bae, Woong and Bellver, Míriam and Bi, Lei and Chen, Hao and Chlebus, Grzegorz and Dam, Erik B. and Dou, Qi and Fu, Chi-Wing and Georgescu, Bogdan and Giró-i-Nieto, Xavier and Gruen, Felix and Han, Xu and Heng, Pheng-Ann and Hesser, Jürgen and Moltz, Jan Hendrik and Igel, Christian and Isensee, Fabian and Jäger, Paul and Jia, Fucang and Kaluva, Krishna Chaitanya and Khened, Mahendra and Kim, Ildoo and Kim, Jae-Hun and Kim, Sungwoong and Kohl, Simon and Konopczynski, Tomasz and Kori, Avinash and Krishnamurthi, Ganapathy and Li, Fan and Li, Hongchao and Li, Junbo and Li, Xiaomeng and Lowengrub, John and Ma, Jun and Maier-Hein, Klaus and Maninis, Kevis-Kokitsi and Meine, Hans and Merhof, Dorit and Pai, Akshay and Perslev, Mathias and Petersen, Jens and Pont-Tuset, Jordi and Qi, Jin and Qi, Xiaojuan and Rippel, Oliver and Roth, Karsten and Sarasua, Ignacio and Schenk, Andrea and Shen, Zengming and Torres, Jordi and Wachinger, Christian and Wang, Chunliang and Weninger, Leon and Wu, Jianrong and Xu, Daguang and Yang, Xiaoping and Yu, Simon Chun-Ho and Yuan, Yading and Yu, Miao and Zhang, Liping and Cardoso, Jorge and Bakas, Spyridon and Braren, Rickmer and Heinemann, Volker and Pal, Christopher and Tang, An and Kadoury, Samuel and Soler, Luc and van Ginneken, Bram and Greenspan, Hayit and Joskowicz, Leo and Menze, Bjoern},
	urldate = {2024-01-03},
	date = {2023-02},
	eprinttype = {arxiv},
	eprint = {1901.04056 [cs]},
}

@article{luo_word_2022,
	title = {{WORD}: A large scale dataset, benchmark and clinical applicable study for abdominal organ segmentation from {CT} image},
	volume = {82},
	issn = {1361-8415},
	url = {https://www.sciencedirect.com/science/article/pii/S1361841522002705},
	doi = {10.1016/j.media.2022.102642},
	shorttitle = {{WORD}},
	pages = {102642},
	journaltitle = {Medical Image Analysis},
	shortjournal = {Medical Image Analysis},
	author = {Luo, Xiangde and Liao, Wenjun and Xiao, Jianghong and Chen, Jieneng and Song, Tao and Zhang, Xiaofan and Li, Kang and Metaxas, Dimitris N. and Wang, Guotai and Zhang, Shaoting},
	urldate = {2024-01-03},
	date = {2022-11-01},
}

@software{monai_consortium_monai_2022,
	title = {{MONAI}: Medical Open Network for {AI}},
	url = {https://zenodo.org/record/7459814},
	shorttitle = {{MONAI}},
	publisher = {Zenodo},
	author = {{MONAI} Consortium},
	urldate = {2023-06-02},
	date = {2022-12-19},
	doi = {10.5281/zenodo.7459814},
}

@article{baldini_addressing_2023,
	title = {Addressing the Contrast Media Recognition Challenge: A Fully Automated Machine Learning Approach for Predicting Contrast Phases in {CT} Imaging},
	volume = {59},
	number = {9},
	journaltitle = {Investigative Radiology},
	author = {Baldini, Giulia and Hosch, René and Schmidt, Cynthia S and Borys, Katarzyna and Kroll, Lennard and Koitka, Sven and Haubold, Patrizia and Pelka, Obioma and Nensa, Felix and Haubold, Johannes},
	date = {2023},
	langid = {english},
}

@article{schockel_developments_2020,
	title = {Developments in X-Ray Contrast Media and the Potential Impact on Computed Tomography},
	volume = {55},
	issn = {1536-0210},
	doi = {10.1097/RLI.0000000000000696},
	pages = {592--597},
	number = {9},
	journaltitle = {Investigative Radiology},
	shortjournal = {Invest Radiol},
	author = {Schöckel, Laura and Jost, Gregor and Seidensticker, Peter and Lengsfeld, Philipp and Palkowitsch, Petra and Pietsch, Hubertus},
	date = {2020-09},
	pmid = {32701620},
}

@incollection{siddiqui_differentiating_2019,
	location = {Cham},
	title = {Differentiating Large from Small Bowel},
	isbn = {978-3-030-26044-6},
	url = {https://doi.org/10.1007/978-3-030-26044-6_78},
	pages = {273--274},
	booktitle = {Essential Radiology Review: A Question and Answer Guide},
	publisher = {Springer International Publishing},
	author = {Siddiqui, Efaza},
	editor = {Eltorai, Adam E. M. and Hyman, Charles H. and Healey, Terrance T.},
	urldate = {2024-01-31},
	date = {2019},
	langid = {english},
	doi = {10.1007/978-3-030-26044-6_78},
}

@article{fu_novel_2018,
	title = {A novel {MRI} segmentation method using {CNN}-based correction network for {MRI}-guided adaptive radiotherapy},
	volume = {45},
	rights = {© 2018 American Association of Physicists in Medicine},
	issn = {2473-4209},
	url = {https://onlinelibrary.wiley.com/doi/abs/10.1002/mp.13221},
	doi = {10.1002/mp.13221},
	pages = {5129--5137},
	number = {11},
	journaltitle = {Medical Physics},
	author = {Fu, Yabo and Mazur, Thomas R. and Wu, Xue and Liu, Shi and Chang, Xiao and Lu, Yonggang and Li, H. Harold and Kim, Hyun and Roach, Michael C. and Henke, Lauren and Yang, Deshan},
	urldate = {2024-01-31},
	date = {2018},
	langid = {english},
	note = {\_eprint: https://onlinelibrary.wiley.com/doi/pdf/10.1002/mp.13221},
}

@article{wang_bowelnet_2023,
	title = {{BowelNet}: Joint Semantic-Geometric Ensemble Learning for Bowel Segmentation From Both Partially and Fully Labeled {CT} Images},
	volume = {42},
	issn = {1558-254X},
	doi = {10.1109/TMI.2022.3225667},
	shorttitle = {{BowelNet}},
	pages = {1225--1236},
	number = {4},
	journaltitle = {{IEEE} transactions on medical imaging},
	shortjournal = {{IEEE} Trans Med Imaging},
	author = {Wang, Chong and Cui, Zhiming and Yang, Junwei and Han, Miaofei and Carneiro, Gustavo and Shen, Dinggang},
	date = {2023-04},
	pmid = {36449590},
}

@online{nvidia_corporation_nvidia_2024,
	title = {{NVIDIA} {RTX} A6000 Powered by Ampere Architecture},
	url = {https://www.nvidia.com/en-us/design-visualization/rtx-a6000/},
	titleaddon = {{NVIDIA}},
	author = {{NVIDIA Corporation}},
	urldate = {2024-01-26},
	date = {2024},
	langid = {english},
}

@article{wasserthal_totalsegmentator_2023,
	title = {{TotalSegmentator}: Robust Segmentation of 104 Anatomic Structures in {CT} Images},
	volume = {5},
	url = {https://pubs.rsna.org/doi/10.1148/ryai.230024},
	doi = {10.1148/ryai.230024},
	shorttitle = {{TotalSegmentator}},
	pages = {e230024},
	number = {5},
	journaltitle = {Radiology: Artificial Intelligence},
	author = {Wasserthal, Jakob and Breit, Hanns-Christian and Meyer, Manfred T. and Pradella, Maurice and Hinck, Daniel and Sauter, Alexander W. and Heye, Tobias and Boll, Daniel T. and Cyriac, Joshy and Yang, Shan and Bach, Michael and Segeroth, Martin},
	urldate = {2023-11-07},
	date = {2023-09},
	note = {Publisher: Radiological Society of North America},
}

@article{yang_autosegmentation_2018,
	title = {Autosegmentation for thoracic radiation treatment planning: A grand challenge at {AAPM} 2017},
	volume = {45},
	issn = {0094-2405},
	url = {https://www.ncbi.nlm.nih.gov/pmc/articles/PMC6714977/},
	doi = {10.1002/mp.13141},
	shorttitle = {Autosegmentation for thoracic radiation treatment planning},
	pages = {4568--4581},
	number = {10},
	journaltitle = {Medical Physics},
	shortjournal = {Med Phys},
	author = {Yang, Jinzhong and Veeraraghavan, Harini and Armato, Samuel G. and Farahani, Keyvan and Kirby, Justin S. and Kalpathy‐Kramer, Jayashree and van Elmpt, Wouter and Dekker, Andre and Han, Xiao and Feng, Xue and Aljabar, Paul and Oliveira, Bruno and van der Heyden, Brent and Zamdborg, Leonid and Lam, Dao and Gooding, Mark and Sharp, Gregory C.},
	urldate = {2023-08-25},
	date = {2018-10},
	pmid = {30144101},
	pmcid = {PMC6714977},
}

@article{clark_cancer_2013,
	title = {The Cancer Imaging Archive ({TCIA}): Maintaining and Operating a Public Information Repository},
	volume = {26},
	issn = {0897-1889, 1618-727X},
	url = {http://link.springer.com/10.1007/s10278-013-9622-7},
	doi = {10.1007/s10278-013-9622-7},
	shorttitle = {The Cancer Imaging Archive ({TCIA})},
	pages = {1045--1057},
	number = {6},
	journaltitle = {Journal of Digital Imaging},
	shortjournal = {J Digit Imaging},
	author = {Clark, Kenneth and Vendt, Bruce and Smith, Kirk and Freymann, John and Kirby, Justin and Koppel, Paul and Moore, Stephen and Phillips, Stanley and Maffitt, David and Pringle, Michael and Tarbox, Lawrence and Prior, Fred},
	urldate = {2023-07-07},
	date = {2013-12},
	langid = {english},
}

@misc{yang_data_2017,
	title = {Data from Lung {CT} Segmentation Challenge 2017 ({LCTSC})},
	rights = {Creative Commons Attribution 3.0 Unported},
	url = {https://www.cancerimagingarchive.net/collection/lctsc/},
	doi = {10.7937/K9/TCIA.2017.3R3FVZ08},
	version = {3},
	publisher = {The Cancer Imaging Archive},
	author = {Yang, Jinzhong and Sharp, Greg and Veeraraghavan, Harini and Van Elmpt, Wouter and Dekker, Andre and Lustberg, Tim and Gooding, Mark},
	editora = {{TCIA Team}},
	editoratype = {collaborator},
	urldate = {2024-01-03},
	date = {2017},
}

@misc{ma_unleashing_2023,
	title = {Unleashing the Strengths of Unlabeled Data in Pan-cancer Abdominal Organ Quantification: the {FLARE}22 Challenge},
	url = {http://arxiv.org/abs/2308.05862},
	doi = {10.48550/arXiv.2308.05862},
	shorttitle = {Unleashing the Strengths of Unlabeled Data in Pan-cancer Abdominal Organ Quantification},
	number = {{arXiv}:2308.05862},
	publisher = {{arXiv}},
	author = {Ma, Jun and Zhang, Yao and Gu, Song and Ge, Cheng and Ma, Shihao and Young, Adamo and Zhu, Cheng and Meng, Kangkang and Yang, Xin and Huang, Ziyan and Zhang, Fan and Liu, Wentao and Pan, {YuanKe} and Huang, Shoujin and Wang, Jiacheng and Sun, Mingze and Xu, Weixin and Jia, Dengqiang and Choi, Jae Won and Alves, Natália and de Wilde, Bram and Koehler, Gregor and Wu, Yajun and Wiesenfarth, Manuel and Zhu, Qiongjie and Dong, Guoqiang and He, Jian and Consortium, the {FLARE} Challenge and Wang, Bo},
	urldate = {2024-01-03},
	date = {2023-08-10},
	eprinttype = {arxiv},
	eprint = {2308.05862 [cs, eess]},
}

@article{armato_iii_lung_2011,
	title = {The Lung Image Database Consortium ({LIDC}) and Image Database Resource Initiative ({IDRI}): A Completed Reference Database of Lung Nodules on {CT} Scans},
	volume = {38},
	issn = {00942405},
	url = {http://doi.wiley.com/10.1118/1.3528204},
	doi = {10.1118/1.3528204},
	shorttitle = {The Lung Image Database Consortium ({LIDC}) and Image Database Resource Initiative ({IDRI})},
	pages = {915--931},
	number = {2},
	journaltitle = {Medical Physics},
	shortjournal = {Med. Phys.},
	author = {Armato {III}, Samuel G. and {McLennan}, Geoffrey and Bidaut, Luc and {McNitt}-Gray, Michael F. and Meyer, Charles R. and Reeves, Anthony P. and Zhao, Binsheng and Aberle, Denise R. and Henschke, Claudia I. and Hoffman, Eric A. and Kazerooni, Ella A. and {MacMahon}, Heber and Van Beek, Edwin J. R. and Yankelevitz, David and Biancardi, Alberto M. and Bland, Peyton H. and Brown, Matthew S. and Engelmann, Roger M. and Laderach, Gary E. and Max, Daniel and Pais, Richard C. and Qing, David P.-Y. and Roberts, Rachael Y. and Smith, Amanda R. and Starkey, Adam and Batra, Poonam and Caligiuri, Philip and Farooqi, Ali and Gladish, Gregory W. and Jude, C. Matilda and Munden, Reginald F. and Petkovska, Iva and Quint, Leslie E. and Schwartz, Lawrence H. and Sundaram, Baskaran and Dodd, Lori E. and Fenimore, Charles and Gur, David and Petrick, Nicholas and Freymann, John and Kirby, Justin and Hughes, Brian and Vande Casteele, Alessi and Gupte, Sangeeta and Sallam, Maha and Heath, Michael D. and Kuhn, Michael H. and Dharaiya, Ekta and Burns, Richard and Fryd, David S. and Salganicoff, Marcos and Anand, Vikram and Shreter, Uri and Vastagh, Stephen and Croft, Barbara Y. and Clarke, Laurence P.},
	urldate = {2023-07-19},
	date = {2011-01-24},
	langid = {english},
}

@misc{tang_automatic_2019,
	title = {Automatic Pulmonary Lobe Segmentation Using Deep Learning},
	url = {http://arxiv.org/abs/1903.09879},
	doi = {10.48550/arXiv.1903.09879},
	number = {{arXiv}:1903.09879},
	publisher = {{arXiv}},
	author = {Tang, Hao and Zhang, Chupeng and Xie, Xiaohui},
	urldate = {2024-01-03},
	date = {2019-04-10},
	eprinttype = {arxiv},
	eprint = {1903.09879 [cs]},
}

@article{haubold_boa_2024,
	title = {{BOA}: A {CT}-Based Body and Organ Analysis for Radiologists at the Point of Care},
	volume = {59},
	url = {http://doi.org/10.1097/RLI.0000000000001040},
	doi = {10.1097/RLI.0000000000001040},
	number = {6},
	journaltitle = {Investigative Radiology},
	author = {Haubold, Johannes and Baldini, Giulia and Parmar, Vicky and Schaarschmidt, Benedikt Michael and Koitka, Sven and Kroll, Lennard and van Landeghem, Natalie and Umutlu, Lale and Forsting, Michael and Nensa, Felix and Hosch, René},
	date = {2024},
	langid = {english},
}

@article{guimaraes_advancements_2023,
	title = {Advancements in non-invasive imaging of atherosclerosis: Future perspectives},
	volume = {0},
	issn = {1933-2874, 1876-4789},
	url = {https://doi.org/10.1016/j.jacl.2023.11.008},
	doi = {10.1016/j.jacl.2023.11.008},
	shorttitle = {Advancements in non-invasive imaging of atherosclerosis},
	number = {0},
	journaltitle = {Journal of Clinical Lipidology},
	shortjournal = {Journal of Clinical Lipidology},
	author = {Guimarães, Joana and Almeida, José de and Mendes, Paulo Lázaro and Ferreira, Maria João and Gonçalves, Lino},
	date = {2023-11-18},
	pmid = {38142178},
	note = {Publisher: Elsevier},
}

@article{osborne-grinter_prevalence_2023,
	title = {Prevalence and clinical implications of coronary artery calcium scoring on non-gated thoracic computed tomography: a systematic review and meta-analysis},
	issn = {1432-1084},
	url = {https://doi.org/10.1007/s00330-023-10439-z},
	doi = {10.1007/s00330-023-10439-z},
	shorttitle = {Prevalence and clinical implications of coronary artery calcium scoring on non-gated thoracic computed tomography},
	journaltitle = {European Radiology},
	shortjournal = {Eur Radiol},
	author = {Osborne-Grinter, Maia and Ali, Adnan and Williams, Michelle C.},
	date = {2023-12-22},
	pmid = {38133672},
}

@article{neubauer_diagnostic_2023,
	title = {Diagnostic Accuracy of Contrast-Enhanced Thoracic Photon-Counting Computed Tomography for Opportunistic Locoregional Staging of Breast Cancer Compared With Digital Mammography: A Prospective Trial},
	issn = {1536-0210},
	url = {https://doi.org/10.1097/RLI.0000000000001051},
	doi = {10.1097/RLI.0000000000001051},
	shorttitle = {Diagnostic Accuracy of Contrast-Enhanced Thoracic Photon-Counting Computed Tomography for Opportunistic Locoregional Staging of Breast Cancer Compared With Digital Mammography},
	journaltitle = {Investigative Radiology},
	shortjournal = {Invest Radiol},
	author = {Neubauer, Jakob and Wilpert, Caroline and Gebler, Oliver and Taran, Florin-Andrei and Pichotka, Martin and Stein, Thomas and Molina-Fuentes, Moisés Felipe and Weiss, Jakob and Juhasz-Böss, Ingolf and Bamberg, Fabian and Windfuhr-Blum, Marisa and Neubauer, Claudia},
	date = {2023-12-01},
	pmid = {38038693},
}

@article{junn_imaging_2021,
	title = {Imaging of Head and Neck Cancer With {CT}, {MRI}, and {US}},
	volume = {51},
	issn = {0001-2998},
	url = {https://www.sciencedirect.com/science/article/pii/S0001299820300763},
	doi = {10.1053/j.semnuclmed.2020.07.005},
	series = {Imaging Options for Head and Neck Cancer},
	pages = {3--12},
	number = {1},
	journaltitle = {Seminars in Nuclear Medicine},
	shortjournal = {Seminars in Nuclear Medicine},
	author = {Junn, Jacqueline C. and Soderlund, Karl A. and Glastonbury, Christine M.},
	urldate = {2023-12-28},
	date = {2021-01-01},
}

@article{hosch_biomarkers_2022,
	title = {Biomarkers extracted by fully automated body composition analysis from chest {CT} correlate with {SARS}-{CoV}-2 outcome severity},
	volume = {12},
	issn = {2045-2322},
	doi = {10.1038/s41598-022-20419-w},
	pages = {16411},
	number = {1},
	journaltitle = {Scientific Reports},
	shortjournal = {Sci Rep},
	author = {Hosch, René and Kattner, Simone and Berger, Marc Moritz and Brenner, Thorsten and Haubold, Johannes and Kleesiek, Jens and Koitka, Sven and Kroll, Lennard and Kureishi, Anisa and Flaschel, Nils and Nensa, Felix},
	date = {2022-09-30},
	pmid = {36180519},
	pmcid = {PMC9524347},
}

@misc{koitka_saros_2023,
	title = {{SAROS} - A large, heterogeneous, and sparsely annotated segmentation dataset on {CT} imaging data ({SAROS})},
	rights = {Creative Commons Attribution 4.0 International},
	url = {https://wiki.cancerimagingarchive.net/x/2wVgCQ},
	doi = {10.25737/SZ96-ZG60},
	version = {1},
	publisher = {The Cancer Imaging Archive},
	author = {Koitka, Sven and Baldini, Giulia and Kroll, Lennard and van Landeghem, Natalie and Haubold, Johannes and Sung Kim, Moon and Kleesiek, Jens and Nensa, Felix and Hosch, Rene},
	urldate = {2023-09-26},
	date = {2023},
}

@article{bodden_incidental_2023,
	title = {Incidental vertebral fracture prediction using neuronal network-based automatic spine segmentation and volumetric bone mineral density extraction from routine clinical {CT} scans},
	volume = {14},
	issn = {1664-2392},
	doi = {10.3389/fendo.2023.1207949},
	pages = {1207949},
	journaltitle = {Frontiers in Endocrinology},
	shortjournal = {Front Endocrinol (Lausanne)},
	author = {Bodden, Jannis and Dieckmeyer, Michael and Sollmann, Nico and Burian, Egon and Rühling, Sebastian and Löffler, Maximilian T. and Sekuboyina, Anjany and El Husseini, Malek and Zimmer, Claus and Kirschke, Jan S. and Baum, Thomas},
	date = {2023},
	pmid = {37529605},
	pmcid = {PMC10390306},
}

@article{jiang_radiomics_2022,
	title = {Radiomics analysis based on lumbar spine {CT} to detect osteoporosis},
	volume = {32},
	issn = {1432-1084},
	doi = {10.1007/s00330-022-08805-4},
	pages = {8019--8026},
	number = {11},
	journaltitle = {European Radiology},
	shortjournal = {Eur Radiol},
	author = {Jiang, Yan-Wei and Xu, Xiong-Jie and Wang, Rui and Chen, Chun-Mei},
	date = {2022-11},
	pmid = {35499565},
	pmcid = {PMC9059457},
}

@article{radiya_performance_2023,
	title = {Performance and clinical applicability of machine learning in liver computed tomography imaging: a systematic review},
	volume = {33},
	issn = {1432-1084},
	doi = {10.1007/s00330-023-09609-w},
	shorttitle = {Performance and clinical applicability of machine learning in liver computed tomography imaging},
	pages = {6689--6717},
	number = {10},
	journaltitle = {European Radiology},
	shortjournal = {Eur Radiol},
	author = {Radiya, Keyur and Joakimsen, Henrik Lykke and Mikalsen, Karl Øyvind and Aahlin, Eirik Kjus and Lindsetmo, Rolv-Ole and Mortensen, Kim Erlend},
	date = {2023-10},
	pmid = {37171491},
	pmcid = {PMC10511359},
}

@article{aromiwura_artificial_2023,
	title = {Artificial intelligence in cardiac computed tomography},
	volume = {81},
	issn = {0033-0620},
	url = {https://www.sciencedirect.com/science/article/pii/S0033062023000920},
	doi = {10.1016/j.pcad.2023.09.001},
	pages = {54--77},
	journaltitle = {Progress in Cardiovascular Diseases},
	shortjournal = {Progress in Cardiovascular Diseases},
	author = {Aromiwura, Afolasayo A. and Settle, Tyler and Umer, Muhammad and Joshi, Jonathan and Shotwell, Matthew and Mattumpuram, Jishanth and Vorla, Mounica and Sztukowska, Maryta and Contractor, Sohail and Amini, Amir and Kalra, Dinesh K.},
	urldate = {2024-01-03},
	date = {2023-11-01},
}

@article{jung_association_2023,
	title = {Association between hypertension and myosteatosis evaluated by abdominal computed tomography},
	volume = {46},
	rights = {2023 The Author(s), under exclusive licence to The Japanese Society of Hypertension},
	issn = {1348-4214},
	url = {https://www.nature.com/articles/s41440-022-01157-y},
	doi = {10.1038/s41440-022-01157-y},
	pages = {845--855},
	number = {4},
	journaltitle = {Hypertension Research},
	shortjournal = {Hypertens Res},
	author = {Jung, Han Na and Cho, Yun Kyung and Kim, Hwi Seung and Kim, Eun Hee and Lee, Min Jung and Lee, Woo Je and Kim, Hong-Kyu and Jung, Chang Hee},
	urldate = {2024-01-03},
	date = {2023-04},
	langid = {english},
	note = {Number: 4
Publisher: Nature Publishing Group},
}

@article{ko_change_2022,
	title = {Change of Computed Tomography-Based Body Composition after Adrenalectomy in Patients with Pheochromocytoma},
	volume = {14},
	issn = {2072-6694},
	doi = {10.3390/cancers14081967},
	pages = {1967},
	number = {8},
	journaltitle = {Cancers},
	shortjournal = {Cancers (Basel)},
	author = {Ko, Yousun and Jeong, Heeryoel and Khang, Seungwoo and Lee, Jeongjin and Kim, Kyung Won and Kim, Beom-Jun},
	date = {2022-04-13},
	pmid = {35454877},
	pmcid = {PMC9024595},
}

@article{nikolov_clinically_2021,
	title = {Clinically Applicable Segmentation of Head and Neck Anatomy for Radiotherapy: Deep Learning Algorithm Development and Validation Study},
	volume = {23},
	issn = {1438-8871},
	doi = {10.2196/26151},
	shorttitle = {Clinically Applicable Segmentation of Head and Neck Anatomy for Radiotherapy},
	pages = {e26151},
	number = {7},
	journaltitle = {Journal of Medical Internet Research},
	shortjournal = {J Med Internet Res},
	author = {Nikolov, Stanislav and Blackwell, Sam and Zverovitch, Alexei and Mendes, Ruheena and Livne, Michelle and De Fauw, Jeffrey and Patel, Yojan and Meyer, Clemens and Askham, Harry and Romera-Paredes, Bernadino and Kelly, Christopher and Karthikesalingam, Alan and Chu, Carlton and Carnell, Dawn and Boon, Cheng and D'Souza, Derek and Moinuddin, Syed Ali and Garie, Bethany and {McQuinlan}, Yasmin and Ireland, Sarah and Hampton, Kiarna and Fuller, Krystle and Montgomery, Hugh and Rees, Geraint and Suleyman, Mustafa and Back, Trevor and Hughes, Cían Owen and Ledsam, Joseph R. and Ronneberger, Olaf},
	date = {2021-07-12},
	pmid = {34255661},
	pmcid = {PMC8314151},
}

@article{alderuccio_quantitative_2023,
	title = {Quantitative {PET}-based biomarkers in lymphoma: getting ready for primetime},
	volume = {20},
	issn = {1759-4782},
	doi = {10.1038/s41571-023-00799-2},
	shorttitle = {Quantitative {PET}-based biomarkers in lymphoma},
	pages = {640--657},
	number = {9},
	journaltitle = {Nature Reviews. Clinical Oncology},
	shortjournal = {Nat Rev Clin Oncol},
	author = {Alderuccio, Juan Pablo and Kuker, Russ A. and Yang, Fei and Moskowitz, Craig H.},
	date = {2023-09},
	pmid = {37460635},
}

@article{khoshpouri_quantitative_2019,
	title = {Quantitative spleen and liver volume changes predict survival of patients with primary sclerosing cholangitis},
	volume = {74},
	issn = {1365-229X},
	doi = {10.1016/j.crad.2019.05.018},
	pages = {734.e13--734.e20},
	number = {9},
	journaltitle = {Clinical Radiology},
	shortjournal = {Clin Radiol},
	author = {Khoshpouri, P. and Hazhirkarzar, B. and Ameli, S. and Pandey, A. and Ghadimi, M. and Rezvani Habibabadi, R. and Aliyari Ghasabeh, M. and Pandey, P. and Shaghaghi, M. and Kamel, I. R.},
	date = {2019-09},
	pmid = {31239109},
}

@article{mason_respiratory_2021,
	title = {Respiratory exacerbations are associated with muscle loss in current and former smokers},
	volume = {76},
	issn = {1468-3296},
	doi = {10.1136/thoraxjnl-2020-215999},
	pages = {554--560},
	number = {6},
	journaltitle = {Thorax},
	shortjournal = {Thorax},
	author = {Mason, Stefanie Elizabeth and Moreta-Martinez, Rafael and Labaki, Wassim W. and Strand, Matthew and Baraghoshi, David and Regan, Elizabeth A. and Bon, Jessica and San Jose Estepar, Ruben and Casaburi, Richard and McDonald, Merry-Lynn N. and Rossiter, Harry and Make, Barry J. and Dransfield, Mark T. and Han, MeiLan K. and Young, Kendra A. and Kinney, Greg and Hokanson, John E. and San Jose Estepar, Raul and Washko, George R. and {COPDGene Investigators} and {COPDGene® Investigators}},
	date = {2021-06},
	pmid = {33574123},
	pmcid = {PMC8222105},
}

@article{el-sappagh_snomed_2018,
	title = {{SNOMED} {CT} standard ontology based on the ontology for general medical science},
	volume = {18},
	issn = {1472-6947},
	url = {https://doi.org/10.1186/s12911-018-0651-5},
	doi = {10.1186/s12911-018-0651-5},
	pages = {76},
	number = {1},
	journaltitle = {{BMC} Medical Informatics and Decision Making},
	shortjournal = {{BMC} Medical Informatics and Decision Making},
	author = {El-Sappagh, Shaker and Franda, Francesco and Ali, Farman and Kwak, Kyung-Sup},
	urldate = {2023-12-13},
	date = {2018-08-31},
}

@article{koitka_saros_2024,
	title = {{SAROS}: A dataset for whole-body region and organ segmentation in {CT} imaging},
	volume = {11},
	rights = {2024 The Author(s)},
	issn = {2052-4463},
	url = {https://www.nature.com/articles/s41597-024-03337-6},
	doi = {10.1038/s41597-024-03337-6},
	shorttitle = {{SAROS}},
	pages = {483},
	number = {1},
	journaltitle = {Scientific Data},
	shortjournal = {Sci Data},
	author = {Koitka, Sven and Baldini, Giulia and Kroll, Lennard and van Landeghem, Natalie and Pollok, Olivia B. and Haubold, Johannes and Pelka, Obioma and Kim, Moon and Kleesiek, Jens and Nensa, Felix and Hosch, René},
	urldate = {2024-07-05},
	date = {2024-05-10},
	langid = {english},
	note = {Publisher: Nature Publishing Group},
}

\newpage

\appendix

\section{Additional Figures}
\label{add-figures}

\begin{figure}[ht!]
\includegraphics[width=\textwidth]{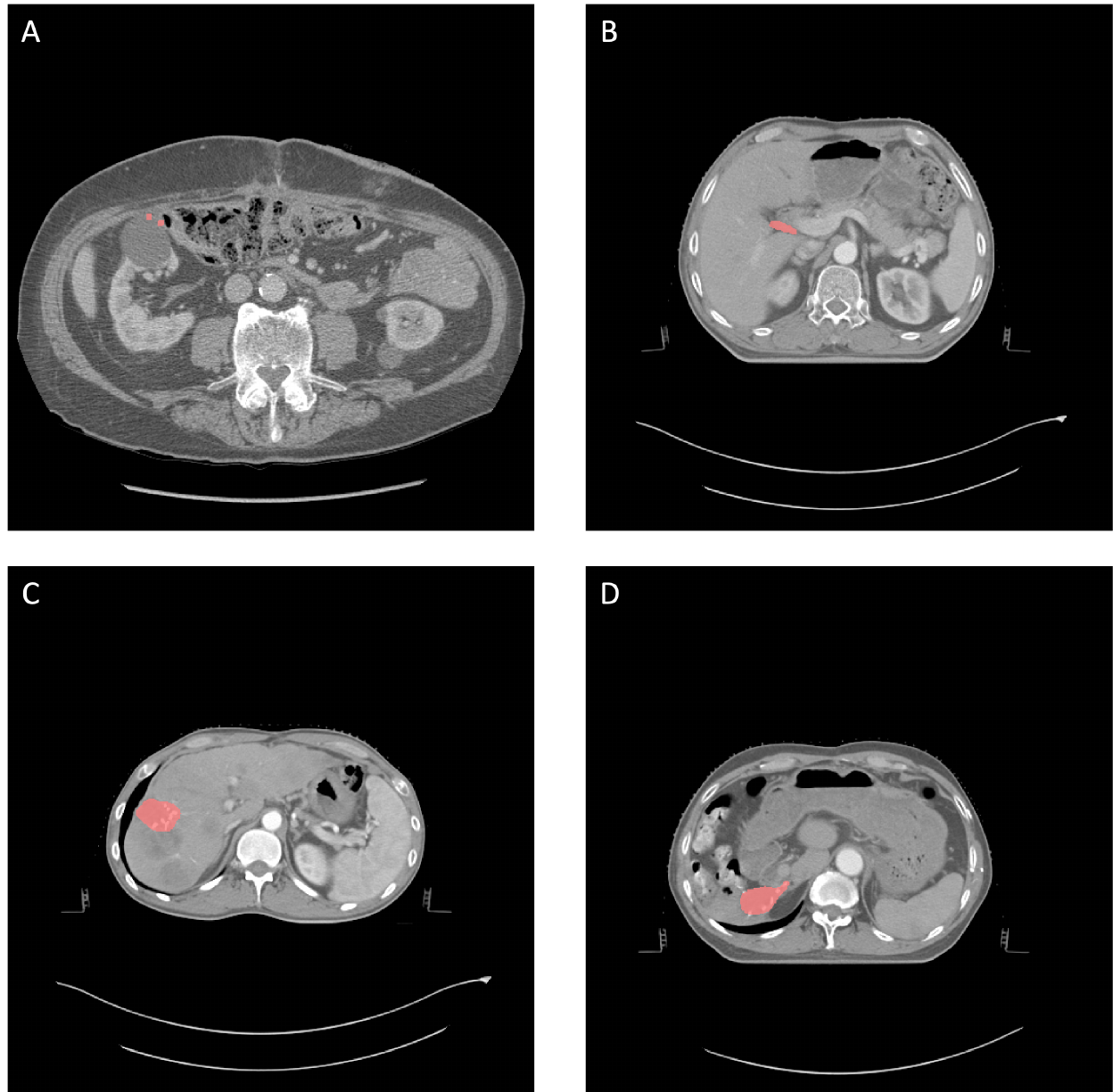}
\captionwithtitle{Gallbladder Segmentations of the WORD dataset}{In A, 5 voxels were marked as gallbladder and binary dilation was performed to be able to visualize them. In B and D, the common bile duct was segmented instead of the gallbladder. In C, a tumor was identified as the gallbladder.}
\label{fig:gallbladder}
\end{figure}

\begin{figure}[ht!]
\includegraphics[width=\textwidth]{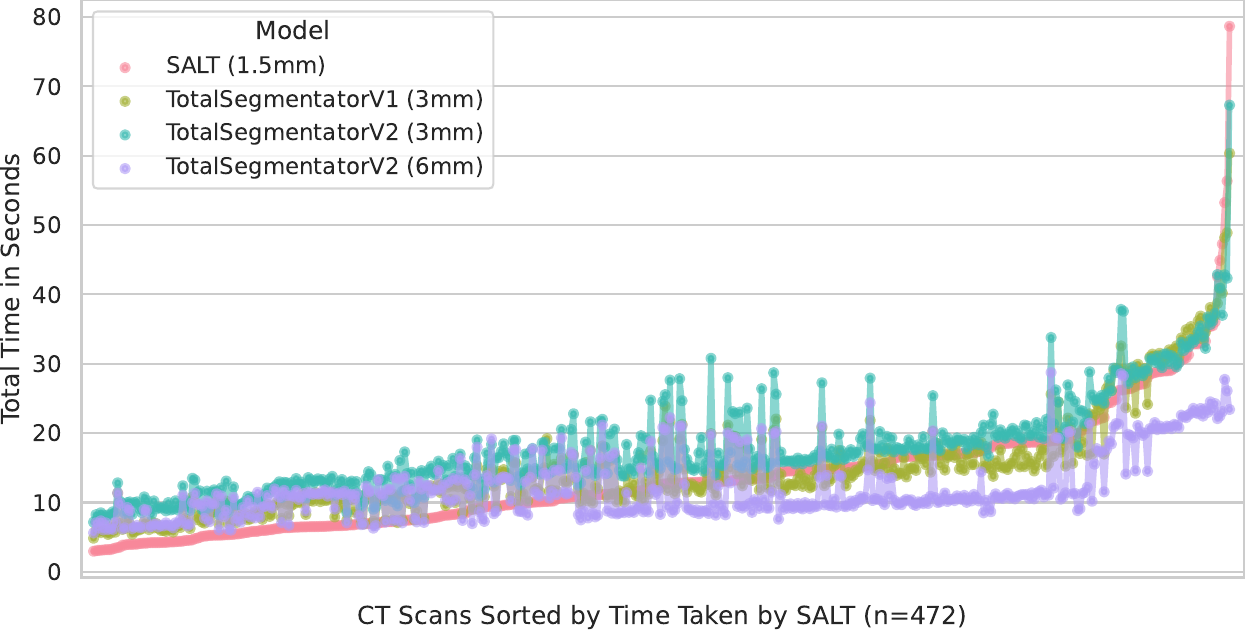}
\captionwithtitle{Comparison between the speed of SALT and the TotalSegmentator}{The fast alternatives of the TotalSegmentator were run on the same set of 472 CT scans from \Cref{tab:datasets}. Version 1 only has a fast alternative that uses one single model and 3mm isotropic spacing, while Version 2 has two alternatives that use 3mm and 6mm isotropic spacing.}
\label{fig:speed2}
\end{figure}

\FloatBarrier
\newpage
\section{Additional Tables}
\label{add-tables}

\begin{table}
\captionwithtitle{Dice scores and Normalized Surface Dice (NSD) for the CT-ORG dataset}{The scores are reported for SALT and Version 2 of the TotalSegmentator (TSV2). The 95\% confidence intervals are reported in brackets.}
\begin{tabularx}{\linewidth}{| c | Y | Y | Y | Y |}
\hline
\textbf{Label} & \textbf{SALT (Dice)} & \textbf{TSV2 (Dice)} & \textbf{SALT (NSD)} & \textbf{TSV2 (NSD)}\\
\hline
Kidneys & 0.872
\par[0.854, 0.884] & 0.927
\par[0.912, 0.94] & 0.914
\par[0.884, 0.937] & 0.951
\par[0.92, 0.975]\\
\hline
Liver & 0.95
\par[0.946, 0.953] & 0.964
\par[0.961, 0.967] & 0.92
\par[0.906, 0.935] & 0.952
\par[0.945, 0.961]\\
\hline
Urinary Bladder & 0.829
\par[0.787, 0.867] & 0.881
\par[0.853, 0.905] & 0.731
\par[0.655, 0.805] & 0.858
\par[0.8, 0.906]\\
\hline
Bones & 0.872
\par[0.863, 0.879] & 0.877
\par[0.861, 0.89] & 0.919
\par[0.912, 0.926] & 0.883
\par[0.866, 0.897]\\
\hline
Brain & 0.486
\par[0, 0.973] & 0.488
\par[0, 0.976] & 0.49
\par[0, 0.981] & 0.492
\par[0, 0.985]\\
\hline
Lungs & 0.962
\par[0.958, 0.966] & 0.97
\par[0.964, 0.975] & 0.968
\par[0.951, 0.982] & 0.971
\par[0.955, 0.984]\\
\hline
\end{tabularx}
\label{tab:ctorg}
\end{table}

\begin{table}
\captionwithtitle{Dice scores and Normalized Surface Dice (NSD) for the FLARE22 dataset}{The scores are reported for SALT and Version 2 of the TotalSegmentator (TSV2). The 95\% confidence intervals are reported in brackets.}
\begin{tabularx}{\linewidth}{| c | Y | Y | Y | Y |}
\hline
\textbf{Label} & \textbf{SALT (Dice)} & \textbf{TSV2 (Dice)} & \textbf{SALT (NSD)} & \textbf{TSV2 (NSD)}\\
\hline
Adrenal Glands & 0.699
\par[0.683, 0.713] & 0.835
\par[0.825, 0.846] & 0.949
\par[0.935, 0.961] & 0.978
\par[0.969, 0.986]\\
\hline
Adrenal Gland (L) & 0.693
\par[0.674, 0.709] & 0.837
\par[0.824, 0.849] & 0.944
\par[0.925, 0.959] & 0.978
\par[0.965, 0.988]\\
\hline
Adrenal Gland (R) & 0.706
\par[0.687, 0.724] & 0.831
\par[0.814, 0.845] & 0.957
\par[0.944, 0.97] & 0.978
\par[0.965, 0.988]\\
\hline
Duodenum & 0.756
\par[0.74, 0.77] & 0.768
\par[0.749, 0.785] & 0.836
\par[0.821, 0.852] & 0.844
\par[0.828, 0.86]\\
\hline
Gallbladder & 0.861
\par[0.844, 0.877] & 0.894
\par[0.878, 0.91] & 0.956
\par[0.942, 0.969] & 0.958
\par[0.941, 0.973]\\
\hline
Kidneys & 0.92
\par[0.913, 0.925] & 0.915
\par[0.899, 0.929] & 0.959
\par[0.948, 0.968] & 0.938
\par[0.922, 0.953]\\
\hline
Kidney (L) & 0.919
\par[0.914, 0.924] & 0.914
\par[0.897, 0.928] & 0.963
\par[0.954, 0.97] & 0.94
\par[0.924, 0.955]\\
\hline
Kidney (R) & 0.921
\par[0.912, 0.928] & 0.921
\par[0.899, 0.936] & 0.956
\par[0.943, 0.967] & 0.939
\par[0.919, 0.955]\\
\hline
Liver & 0.952
\par[0.95, 0.954] & 0.972
\par[0.971, 0.973] & 0.966
\par[0.96, 0.97] & 0.977
\par[0.975, 0.979]\\
\hline
Pancreas & 0.843
\par[0.83, 0.856] & 0.833
\par[0.815, 0.85] & 0.934
\par[0.92, 0.946] & 0.929
\par[0.91, 0.945]\\
\hline
Spleen & 0.94
\par[0.935, 0.943] & 0.974
\par[0.972, 0.975] & 0.972
\par[0.966, 0.978] & 0.999
\par[0.998, 0.999]\\
\hline
Stomach & 0.93
\par[0.925, 0.935] & 0.954
\par[0.949, 0.958] & 0.974
\par[0.967, 0.979] & 0.982
\par[0.976, 0.987]\\
\hline
Aorta & 0.89
\par[0.882, 0.896] & 0.936
\par[0.926, 0.945] & 0.951
\par[0.937, 0.962] & 0.958
\par[0.94, 0.973]\\
\hline
Vena Cava Inferior & 0.859
\par[0.854, 0.865] & 0.912
\par[0.905, 0.918] & 0.942
\par[0.934, 0.949] & 0.954
\par[0.946, 0.96]\\
\hline
\end{tabularx}
\label{tab:flare22}
\end{table}

\begin{table}
\captionwithtitle{Dice scores and Normalized Surface Dice (NSD) for the LCTSC dataset}{The scores are reported for SALT and Version 2 of the TotalSegmentator (TSV2). The 95\% confidence intervals are reported in brackets. (L) = Left, (R) = Right.}
\begin{tabularx}{\linewidth}{| c | Y | Y | Y | Y |}
\hline
\textbf{Label} & \textbf{SALT (Dice)} & \textbf{TSV2 (Dice)} & \textbf{SALT (NSD)} & \textbf{TSV2 (NSD)}\\
\hline
Spinal Cord & 0.84
\par[0.833, 0.845] & 0.881
\par[0.874, 0.886] & 0.969
\par[0.963, 0.975] & 0.977
\par[0.971, 0.981]\\
\hline
Lungs & 0.941
\par[0.934, 0.947] & 0.959
\par[0.953, 0.965] & 0.894
\par[0.877, 0.909] & 0.937
\par[0.925, 0.949]\\
\hline
Lung (L) & 0.922
\par[0.9, 0.94] & 0.948
\par[0.933, 0.96] & 0.878
\par[0.849, 0.904] & 0.929
\par[0.908, 0.948]\\
\hline
Lung (R) & 0.946
\par[0.937, 0.952] & 0.961
\par[0.953, 0.967] & 0.896
\par[0.877, 0.913] & 0.938
\par[0.921, 0.952]\\
\hline
Pericardium & 0.894
\par[0.887, 0.901] & / & 0.792
\par[0.769, 0.815] & /\\
\hline
\end{tabularx}
\label{tab:lctsc}
\end{table}

\begin{table}
\captionwithtitle{Dice scores and Normalized Surface Dice (NSD) for the LUNA16 dataset}{The scores are reported for SALT and Version 2 of the TotalSegmentator (TSV2). The 95\% confidence intervals are reported in brackets. (L) = Left, (R) = Right.}
\begin{tabularx}{\linewidth}{| c | Y | Y | Y | Y |}
\hline
\textbf{Label} & \textbf{SALT (Dice)} & \textbf{TSV2 (Dice)} & \textbf{SALT (NSD)} & \textbf{TSV2 (NSD)}\\
\hline
Lungs & 0.961
\par[0.956, 0.964] & 0.983
\par[0.981, 0.984] & 0.951
\par[0.935, 0.962] & 0.979
\par[0.976, 0.982]\\
\hline
Lung (L) & 0.956
\par[0.949, 0.961] & 0.982
\par[0.979, 0.984] & 0.945
\par[0.926, 0.958] & 0.981
\par[0.977, 0.983]\\
\hline
Lung Lower Lobe (L) & 0.931
\par[0.917, 0.94] & 0.961
\par[0.956, 0.966] & 0.905
\par[0.875, 0.927] & 0.953
\par[0.941, 0.964]\\
\hline
Lung Upper Lobe (L) & 0.948
\par[0.943, 0.952] & 0.969
\par[0.966, 0.971] & 0.943
\par[0.93, 0.953] & 0.962
\par[0.952, 0.97]\\
\hline
Lung (R) & 0.964
\par[0.96, 0.967] & 0.984
\par[0.982, 0.985] & 0.956
\par[0.943, 0.965] & 0.978
\par[0.975, 0.981]\\
\hline
Lung Lower Lobe (R) & 0.93
\par[0.918, 0.938] & 0.955
\par[0.948, 0.962] & 0.894
\par[0.871, 0.912] & 0.926
\par[0.913, 0.939]\\
\hline
Lung Middle Lobe (R) & 0.835
\par[0.787, 0.873] & 0.848
\par[0.8, 0.885] & 0.794
\par[0.75, 0.835] & 0.806
\par[0.763, 0.842]\\
\hline
Lung Upper Lobe (R) & 0.916
\par[0.894, 0.931] & 0.933
\par[0.911, 0.947] & 0.874
\par[0.845, 0.898] & 0.888
\par[0.862, 0.911]\\
\hline
\end{tabularx}
\label{tab:luna16}
\end{table}

\begin{table}
\captionwithtitle{Dice scores and Normalized Surface Dice (NSD) for the SAROS dataset}{The scores are reported for SALT and Version 2 of the TotalSegmentator (TSV2). The 95\% confidence intervals are reported in brackets.}
\begin{tabularx}{\linewidth}{| c | Y | Y | Y | Y |}
\hline
\textbf{Label} & \textbf{SALT (Dice)} & \textbf{TSV2 (Dice)} & \textbf{SALT (NSD)} & \textbf{TSV2 (NSD)}\\
\hline
Abdominal Cavity & 0.98
\par[0.979, 0.981] & / & 0.995
\par[0.994, 0.995] & /\\
\hline
Bones & 0.911
\par[0.908, 0.914] & 0.805
\par[0.8, 0.811] & 0.991
\par[0.99, 0.992] & 0.942
\par[0.938, 0.945]\\
\hline
Brain & 0.758
\par[0.656, 0.858] & 0.939
\par[0.886, 0.974] & 0.794
\par[0.691, 0.896] & 0.964
\par[0.91, 0.996]\\
\hline
Muscles & 0.931
\par[0.928, 0.934] & / & 0.989
\par[0.987, 0.991] & /\\
\hline
Spinal Cord & 0.849
\par[0.845, 0.852] & 0.878
\par[0.872, 0.885] & 0.975
\par[0.972, 0.978] & 0.964
\par[0.958, 0.97]\\
\hline
Subcutaneous Tissue & 0.937
\par[0.932, 0.942] & / & 0.991
\par[0.989, 0.993] & /\\
\hline
Thoracic Cavity & 0.972
\par[0.969, 0.975] & / & 0.992
\par[0.99, 0.994] & /\\
\hline
Mediastinum & 0.95
\par[0.944, 0.955] & / & 0.981
\par[0.976, 0.984] & /\\
\hline
Pericardium & 0.952
\par[0.949, 0.955] & / & 0.982
\par[0.979, 0.984] & /\\
\hline
\end{tabularx}
\label{tab:saros}
\end{table}

\begin{table}
\captionwithtitle{Dice scores and Normalized Surface Dice (NSD) for the WORD dataset}{The scores are reported for SALT and Version 2 of the TotalSegmentator (TSV2). The 95\% confidence intervals are reported in brackets. (L) = Left, (R) = Right.}

\begin{tabularx}{\linewidth}{| c | Y | Y | Y | Y |}
\hline
\textbf{Label} & \textbf{SALT (Dice)} & \textbf{TSV2 (Dice)} & \textbf{SALT (NSD)} & \textbf{TSV2 (NSD)}\\
\hline
Adrenal Glands & 0.65
\par[0.631, 0.667] & 0.624
\par[0.603, 0.643] & 0.887
\par[0.868, 0.903] & 0.856
\par[0.833, 0.875]\\
\hline
Colon & 0.808
\par[0.796, 0.819] & 0.773
\par[0.76, 0.785] & 0.828
\par[0.814, 0.841] & 0.784
\par[0.768, 0.798]\\
\hline
Duodenum & 0.642
\par[0.619, 0.662] & 0.623
\par[0.598, 0.644] & 0.741
\par[0.719, 0.761] & 0.718
\par[0.692, 0.74]\\
\hline
Kidneys & 0.923
\par[0.917, 0.926] & 0.925
\par[0.921, 0.928] & 0.97
\par[0.963, 0.974] & 0.973
\par[0.968, 0.977]\\
\hline
Kidney (L) & 0.922
\par[0.914, 0.928] & 0.922
\par[0.918, 0.926] & 0.971
\par[0.961, 0.978] & 0.969
\par[0.963, 0.974]\\
\hline
Kidney (R) & 0.922
\par[0.92, 0.925] & 0.927
\par[0.923, 0.931] & 0.969
\par[0.965, 0.972] & 0.976
\par[0.971, 0.98]\\
\hline
Liver & 0.951
\par[0.949, 0.953] & 0.956
\par[0.954, 0.957] & 0.952
\par[0.946, 0.958] & 0.96
\par[0.956, 0.963]\\
\hline
Pancreas & 0.802
\par[0.79, 0.812] & 0.791
\par[0.782, 0.8] & 0.922
\par[0.914, 0.93] & 0.911
\par[0.903, 0.918]\\
\hline
Small Bowel & 0.821
\par[0.81, 0.83] & 0.797
\par[0.784, 0.81] & 0.867
\par[0.855, 0.878] & 0.847
\par[0.831, 0.863]\\
\hline
Spleen & 0.927
\par[0.924, 0.93] & 0.939
\par[0.937, 0.942] & 0.962
\par[0.956, 0.968] & 0.986
\par[0.983, 0.989]\\
\hline
Stomach & 0.9
\par[0.89, 0.907] & 0.904
\par[0.898, 0.91] & 0.925
\par[0.914, 0.935] & 0.93
\par[0.922, 0.938]\\
\hline
Urinary Bladder & 0.864
\par[0.843, 0.882] & 0.903
\par[0.884, 0.918] & 0.914
\par[0.894, 0.931] & 0.96
\par[0.945, 0.971]\\
\hline
\end{tabularx}
\label{tab:word}
\end{table}

\begin{table}
\captionwithtitle{Evaluation of the time taken by the model at inference time}{The inference and total time are reported in seconds, and all values are given as mean ± standard deviation. The times and the number of slices were stored after preprocessing the CT scan to a spacing of (1.5, 1.5, 1.5).}
{\footnotesize
\begin{tabularx}{\linewidth}{| Y | c | c | c | c | c | c |}
\hline
& \textbf{CT-ORG} & \textbf{FLARE22} & \textbf{LCTSC} & \textbf{LUNA16} & \textbf{SAROS} & \textbf{WORD}\\
\hline
Inference Time & 8.87 ± 11.88 & 2.84 ± 1.19 & 8.89 ± 2.84 & 2.74 ± 1.03 & 8.86 ± 7.68 & 10.25 ± 2.83\\
\hline
Total Time & 17.73 ± 14.77 & 4.95 ± 1.46 & 13.21 ± 3.56 & 6.96 ± 2.01 & 15.95 ± 11.63 & 15.93 ± 3.03\\
\hline
Number Slices & 396.19 ± 187.16 & 167.64 ± 39.19 & 273.76 ± 37.0 & 208.68 ± 26.94 & 359.05 ± 183.77 & 385.44 ± 61.75\\
\hline
Seconds per Slice & 0.04 ± 0.01 & 0.02 ± 0.008 & 0.04 ± 0.01 & 0.03 ± 0.009 & 0.04 ± 0.01 & 0.04 ± 0.003\\
\hline
\end{tabularx}
}
\label{tab:time}
\end{table}

\begin{table}
\captionwithtitle{Results of the statistical tests for the Dice score and the Normalized Surface Dice (NSD) across the datasets}{For organs present in multiple datasets, the results across these datasets were compared using statistical tests. The comparisons were conducted either using the Mann-Whitney U test (for labels belonging to two datasets) or with the Kruskal-Wallis test with Dunn’s post-hoc multiple comparison test adjustment (for more than two datasets). Moreover, the p-values were adjusted for multiple comparisons using Bonferroni’s method. A star (*) denotes statistically significant differences in results between the reported datasets for a particular organ label. In all cases, p-values less than or equal to 0.05 were considered significant.}
\begin{tabularx}{\linewidth}{| c | Y | Y |}
\hline
& \textbf{Dice} & \textbf{NSD}\\
\hline
Adrenal Glands & FLARE22 vs. WORD: 0.0433* & FLARE22 vs. WORD: 2.6896e-06*\\
\hline
Bones & CT-ORG vs. SAROS: 2.1364e-09* & CT-ORG vs. SAROS: 4.2888e-12*\\
\hline
Brain & CT-ORG vs. SAROS: 1.0 & CT-ORG vs. SAROS: 1.0\\
\hline
Duodenum & FLARE22 vs. WORD: 1.3189e-09* & FLARE22 vs. WORD: 1.5321e-06*\\
\hline
Kidney Left & FLARE22 vs. WORD: 0.2095 & FLARE22 vs. WORD: 0.0042*\\
\hline
Kidney Right & FLARE22 vs. WORD: 1.0 & FLARE22 vs. WORD: 0.8368\\
\hline
Kidneys & CT-ORG vs. FLARE22: 1.7019e-06*\par
CT-ORG vs. WORD: 3.6064e-10*\par
FLARE22 vs. WORD: 1.0 & CT-ORG vs. FLARE22: 0.0050*\par
CT-ORG vs. WORD: 2.5901e-09*\par
FLARE22 vs. WORD: 0.1195\\
\hline
Liver & CT-ORG vs. FLARE22: 1.0\par
CT-ORG vs. WORD: 1.0\par
FLARE22 vs. WORD: 1.0 & CT-ORG vs. FLARE22: 2.4277e-05*\par
CT-ORG vs. WORD: 0.0029*\par
FLARE22 vs. WORD: 1.0\\
\hline
Lung Left & LCTSC vs. LUNA16: 1.0141e-05* & LCTSC vs. LUNA16: 6.1510e-06*\\
\hline
Lung Right & LCTSC vs. LUNA16: 1.9292e-05* & LCTSC vs. LUNA16: 1.3627e-08*\\
\hline
Lungs & CT-ORG vs. LCTSC: 0.0020*\par
CT-ORG vs. LUNA16: 1.0\par
LCTSC vs. LUNA16: 2.2457e-06* & CT-ORG vs. LCTSC: 3.8623e-08*\par
CT-ORG vs. LUNA16: 1.0\par
LCTSC vs. LUNA16: 1.1547e-06*\\
\hline
Pancreas & FLARE22 vs. WORD: 0.0006* & FLARE22 vs. WORD: 1.0\\
\hline
Pericardium & LCTSC vs. SAROS: 1.1399e-23* & LCTSC vs. SAROS: 1.5610e-27*\\
\hline
Spinal Cord & LCTSC vs. SAROS: 0.5081 & LCTSC vs. SAROS: 1.0\\
\hline
Spleen & FLARE22 vs. WORD: 0.0001* & FLARE22 vs. WORD: 1.0\\
\hline
Stomach & FLARE22 vs. WORD: 3.3810e-06* & FLARE22 vs. WORD: 2.4462e-09*\\
\hline
Urinary Bladder & CT-ORG vs. WORD: 0.1782 & CT-ORG vs. WORD: 5.6032e-06*\\
\hline
\end{tabularx}\label{tab:statistics}
\end{table}

\newpage
\FloatBarrier

\section{Implementation Details}
\label{implementation-details}

\begin{figure}[!ht]
\centering
\includegraphics[width=0.5\textwidth]{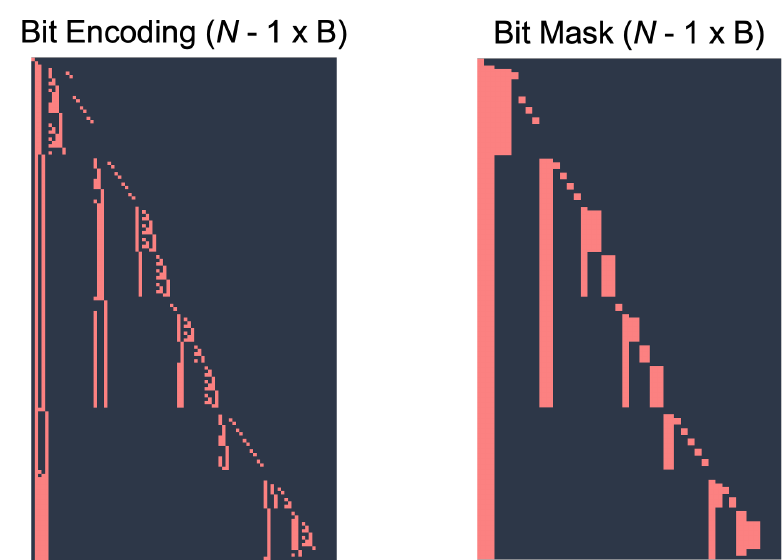}
\captionwithtitle{Bitwise representation of the class hierarchy}{Both the bitwise encoding and the bitwise mask have size $N \times B$, representing the number of classes ($N$) and the number of bytes ($B$) needed to uniquely identify each node.}
\label{fig:bits}
\end{figure}

To evaluate the model's performance during training, a different Dice score formulation was devised. This approach uses the tree structure to create an encoding for each node, thereby eliminating the need to merge masks for evaluation purposes. The standard definition of Dice score can be computed using true positives (TP), false positives (FP), and false negatives (FN):
\begin{alignat*}{2}
\text{Dice} = \frac{2\cdot \text{TP}}{\text{TP}+\text{FP}+\text{FN}}.
\end{alignat*}
Let $y_c$ and $\hat{y}_c$ be binary vectors representing the ground truth and the prediction for a specific class $c$. Then TP, FP, and FN can be computed for the same class $c$ using simple logical operations such as a logical and ($\land$) and logical not ($\neg$) and by computing the bit summation over each voxel $v$ in the voxels set $V$:
\begin{alignat*}{2}
\text{TP}_c&=\sum_{v \in V}\hat{y}^{(v)}_c\land y^{(v)}_c\\
\text{FP}_c&=\sum_{v \in V}\hat{y}^{(v)}_c\land \neg y^{(v)}_c\\
\text{FN}_c&=\sum_{v \in V}\neg\hat{y}^{(v)}_c\land y^{(v)}_c
\end{alignat*}
To effectively assess the hierarchical relationships among classes, each class is assigned a bitwise encoding. The size of the encoding depends on the amount of bytes needed for uniquely encoding the node relationships. This means that the size of the encoding is $N \times B$, where $B$ is the number of bytes. The bitwise encoding at position $c$ of the matrix has size $B$ and it is a signature that represents which nodes are traversed to get to node $c$, i.e., all the parents of the node. Additionally, a bitwise mask of size $N \times B$ is also computed to mask the relevant bits of the encoding that correspond to the parents and the siblings of each node $c$. An example of the bitwise encoding and mask is shown in \Cref{fig:bits}. For bitwise signatures with no more than 64 bits of encoding, native data types from the numpy package57 can be employed. In the other cases, the encoding can be split into individual bytes, and it requires an additional boolean logical and reduction operation to check for a binary match.

Using these two matrices, it is possible to compute the Dice for voxel $v$ and class $c$ by converting the ground truth ($y$) and the prediction ($\hat{y}$) to a binary representation that uses these encodings. Let $v$ be a voxel of the ground truth $y$ or of the prediction $\hat{y}$, and let $c$ be the class for which the Dice is computed. To compute the binary representation $\text{bin}(v, c)$, the voxel $v$ is first converted to its bit encoding, and then it is masked using the bitwise mask of class $c$:
$\text{bin}(v, c) = \text{encoding}\left(v\right)\land\text{mask}\left(c\right).$
The result is a binary representation of the hierarchical affiliation of the voxel $v$ to the class $c$. In practice, this is done for a set of voxels $V$ at the same time, resulting in a two-dimensional array of binary representations of each voxel $v \in V$. The binary representation is then compared to the binary encoding $\text{encoding}\left(c\right)$ of class $c$ with an equality operator ($\text{equals}$):
\begin{alignat*}{2}
\text{equals}(x, y) = \begin{cases}
    \begin{array}{ll}1 & \mbox{if } x = y \\
    0 & \mbox{otherwise}\end{array},
\end{cases}
\end{alignat*}
and the ground truth and the prediction are redefined as:
\begin{alignat*}{2}
y'_c &= \text{equals}(\text{bin}(y_c, c), \text{encoding}\left(c\right)),\\
\hat{y}'_c &= \text{equals}(\text{bin}(\hat{y}_c, c),\text{encoding}\left(c\right)).
\end{alignat*}
Now, $y'_c$ and $\hat{y}'_c$ can be used to compute the Dice score with the true positive, false negative, and false positive formulas shown above.

\end{document}